%% file: main.tex
\documentclass[pageno]{jpaper}

\usepackage[normalem]{ulem}
\usepackage{listings}
\usepackage{amsmath}
\usepackage[noend,linesnumbered,ruled,vlined]{algorithm2e}
\usepackage{physics}
\usepackage{amsmath}
\usepackage{multirow}
\usepackage{enumitem}
\usepackage{amssymb}
\newcommand{\ignore}[1]{}
\usepackage{adjustbox}
\usepackage[normalem]{ulem}
\usepackage[flushleft]{threeparttable}
\usepackage{dblfloatfix}

\usepackage{graphicx}
\usepackage{subfig}

\SetCommentSty{mycommfont}

\def\BibTeX{{\rm B\kern-.05em{\sc i\kern-.025em b}\kern-.08em
    T\kern-.1667em\lower.7ex\hbox{E}\kern-.125emX}}

\usepackage{url}

\usepackage[noadjust]{cite}
\usepackage{tikz}

\def\BibTeX{{\rm B\kern-.05em{\sc i\kern-.025em b}\kern-.08em
    T\kern-.1667em\lower.7ex\hbox{E}\kern-.125emX}}
\definecolor{aliceblue}{rgb}{0.94, 0.97, 1.0}
\usepackage{xcolor}
\usepackage{tcolorbox}
\tcbset{width=1\columnwidth,boxrule=1pt,colback=aliceblue,arc=1pt,left=2pt,right=2pt,boxsep=0.5pt}

\usepackage{authblk}
\begin{document}

\author[1,2,3]{Poulami Das}
\author[1]{Aditya Locharla}
\author[1]{Cody Jones}
\affil[1]{Google Quantum AI, Google Research, Mountain View, CA, USA}
\affil[2]{Department of Electrical and Computer Engineering, Georgia Institute of Technology, Atlanta, GA, USA}
\affil[3]{Email: poulami@gatech.edu}
\title{
LILLIPUT: A Lightweight Low-Latency Lookup-Table Based Decoder for Near-term Quantum Error Correction}

\date{}
\maketitle

\thispagestyle{empty}

\input{./sections/section0abstract}
\input{./sections/section1introduction}

\input{./sections/section2backgroundandmotivation}
\input{./sections/section3methodology}
\input{./sections/section4design}
\input{./sections/section5lut}
\input{./sections/section6results}

\input{./sections/section7clut}

\input{./sections/section8relatedwork}
\input{./sections/section9conclusion}

\bibliographystyle{plain}
\bibliography{references}

\end{document}

%% file: sections/section0abstract.tex
\begin{abstract}
The error rates of quantum devices are orders of magnitude higher than what is needed to run most quantum applications. To close this gap, Quantum Error Correction (QEC) encodes logical qubits and distributes information using several physical qubits. By periodically executing a syndrome extraction circuit on the logical qubits, information about errors (called syndrome) is extracted while running programs. A decoder uses these syndromes to identify and correct errors in real time, which is required to use feedback implemented in quantum algorithms~\cite{jones2012layered}. Unfortunately, software decoders are slow and hardware decoders are fast but less accurate. Thus, almost all QEC studies so far have relied on offline decoding. 

To enable real-time decoding in near-term QEC, we propose LILLIPUT-- a \underline{Li}ghtweight \underline{L}ow Latency \underline{L}ook-\underline{U}p \underline{T}able decoder. LILLIPUT consists of two parts-- First, it translates syndromes into error detection events that index into a Look-Up Table (LUT) whose entry provides the error information in real-time. Second, it programs the LUTs with error assignments for all possible error events by running a software decoder offline. LILLIPUT tolerates an error on any operation in the quantum hardware, including gates and measurement, and the number of tolerated errors grows with the size of the code. It needs <7\% logic on off-the-shelf FPGAs that allows it to be easily integrated alongside the control and readout circuits in existing systems~\cite{googlesupremacy}. LILLIPUT incurs a latency of few nanoseconds and enables real-time decoding. We also propose Compressed LUTs (CLUTs) to reduce the memory needed by LILLIPUT. By exploiting the fact that not all error events are equally likely and only storing data for the most probable error events, CLUTs reduce the memory needed by up-to 107x (from 148 MB to 1.38 MB) without degrading accuracy.

\end{abstract}

%% file: sections/section1introduction.tex
\section{Introduction}
\label{sec:intro}

Quantum computers promise substantial speedup over conventional machines for many important applications~\cite{lloyd1996universal,shor1999polynomial}. Unfortunately, high error-rates of quantum devices (about 1\% on existing hardware~\cite{sycamoredatasheet}) limit us from running these applications as they require lower error-rates (below $10^{-10}$)~\cite{lemieux2021resource,lee2020even,gidney2021factor,kivlichan2020improved}. To bridge this gap, quantum information must be protected by using Quantum Error Correction (QEC). QEC codes encode a logical qubit by distributing information over many physical qubits~\cite{shor1996fault,aharonov1999fault,aliferis2005quantum,gottesman2010introduction,fowler2012surface}. With increasing redundancy of the QEC code, the logical error-rate reduces exponentially if the physical error-rate is below a \textit{threshold}~\cite{aharonov1999fault}. Thus, by controlling the redundancy, QEC enables us to achieve the error-rate needed to run a particular application.

\begin{figure*}[htb]

  \includegraphics[width=\linewidth]{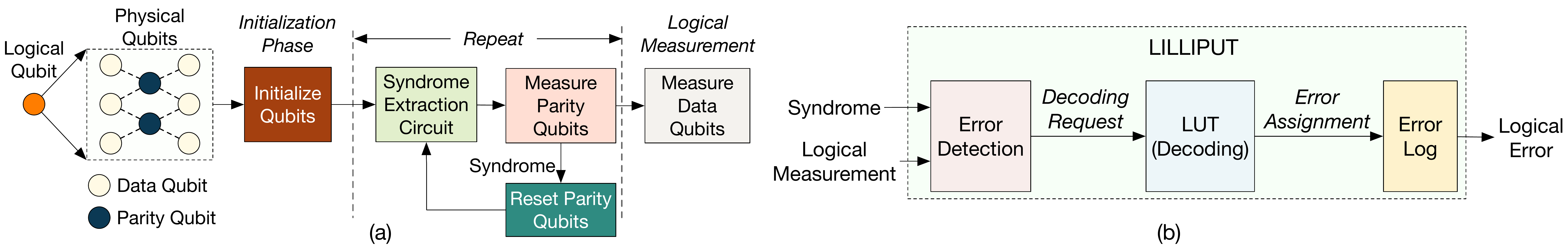}
\caption{(a) In QEC, a logical qubit is encoded using a set of \textit{data} and \textit{parity} qubits. The parity qubits repeatedly extract the syndrome between qubit initialization and logical measurement. (b) LILLIPUT takes the syndrome and logical measurement outcomes as inputs, detects errors, obtains the error assignments for each data qubit from a Look-Up Table (LUT), and tracks an error log every QEC cycle. Finally, it computes the logical error by comparing the logical measurement and the error log.}
\label{fig:intro}
\end{figure*}

QEC consist of both quantum and classical counterparts. On the quantum side, a logical qubit encodes quantum information in the combined state of multiple \textit{data} qubits and uses \textit{parity} qubits interspersed between them to gather information about errors, as shown in Figure~\ref{fig:intro}(a). Each parity qubit periodically extracts the parity information of a subset of data qubits by executing a \textit{syndrome extraction circuit} and projects the errors encountered by the data qubits into discrete errors. The process is repeated from the time qubits are initialized and until the data qubits are measured (called \textit{logical measurement}). Each iteration of syndrome extraction is termed as a \textit{QEC cycle} or \textit{round} and the measurement outcome of all the parity qubits is called a \textit{syndrome}. On the classical side, a \textit{decoder} uses the syndromes to detect errors and determine the optimal correction for the data qubits. The logical error-rate depends on the physical error-rate as well as the performance of the decoder. High physical error rates make QEC codes ineffective~\cite{aharonov1999fault}. Similarly, if a decoder is inaccurate or fails to determine the correction in real-time, errors can accumulate. \textit{Real-time decoding} refers to the detection and correction of errors dynamically as syndromes are received in each QEC cycle and before the arrival of the syndrome in the next cycle. Inaccurate decoding or failure to decode in real-time may lead to logical failure during program execution. Therefore, accurate and real-time decoders are essential for QEC. 

In recent years, several preliminary QEC experiments involving repetition codes and Bacon-Shor codes have been successfully demonstrated~\cite{cory1998experimental,schindler2011experimental,moussa2011demonstration,zhang2011experimental,reed2012realization,waldherr2014quantum,riste2015detecting,cramer2016repeated,wootton2018repetition,wootton2020benchmarking,chen2021exponential,egan2020fault,luo2020quantum}. However, reaching low logical error rates requires implementation of more efficient QEC codes such as surface codes~\cite{kitaev2003fault}, and recent studies~\cite{chen2021exponential,andersen2020repeated,bell2014experimental,marques2021logical} have taken a step in this direction. Unfortunately, QEC studies so far have mainly resorted to offline decoding because most software decoders are slow~\cite{fowler2017QEC_talk}. 
Alternately hardware decoders are faster but have poor accuracy. Additionally, hardware decoders require custom design~\cite{das2020scalable} or superconducting devices~\cite{nisqplus,ueno2021qecool} and therefore, building them in near-term requires significant engineering effort, are expensive, and sometimes impractical due to technology limitations (superconducting decoders for example).  With improving device quality and size of quantum systems, QEC studies involving small surface codes that span multiple QEC cycles and use real-time decoding represent the next significant milestone in QEC. Consequently, there is a growing demand and need for accurate, fast, and low-cost decoding solutions in near-term QEC. To address this challenge, in this paper, we propose \textit{LILLIPUT}-- a Lightweight Low Latency Look-Up Table decoder for small surface codes.

Our proposed design LILLIPUT is an end-to-end system that directly interfaces with the qubit readout circuits, detects and corrects errors in real-time, and computes the logical error. LILLIPUT performs three key steps in hardware, shown in Figure~\ref{fig:intro}(b)-- \textbf{(1)} translates the syndromes every QEC cycle into \textit{error detection events}, \textbf{(2)} assigns errors to each data qubit from a Look-Up Table (LUT) and maintains an error log in real-time, and \textbf{(3)} computes the logical error by comparing the logical measurement outcomes with the most up-to-date error log. Additionally, LILLIPUT programs the LUT by running a software decoder offline for all possible error events. 
 
For high accuracy, a decoder must correct (a) errors that accumulate on data qubits, (b) gate and (c) measurement errors in the syndrome extraction circuit, and (d) errors on data qubit measurements during the logical-measurement stage. LILLIPUT tackles all these errors by decoding multiple rounds of syndrome and by converting the logical measurement data into an appropriate syndrome in the last cycle. The accuracy and complexity of a decoder depends on the \textit{decoder configuration}, a combination of the QEC code redundancy and the number of syndrome rounds used in decoding. As LILLIPUT is fully modular and reconfigurable, it can be implemented across a wide range of decoder configurations. LILLIPUT programs the LUTs offline using the software Minimum Weight Perfect Matching (MWPM) decoder~\cite{fowler2012MWPM,fowler2013minimum}, widely used for its combination of high accuracy and polynomial time complexity. The MWPM decoder may produce multiple possible error assignments for a single error event. For greater accuracy, LILLIPUT considers the error model of the device (we use Google Sycamore~\cite{sycamoredatasheet}) and selects the most probable error. As LILLIPUT is reconfigurable, it can be adapted to other decoding algorithms, device error models, and QEC codes.

Instead of determining errors using hardware or software at run time, LILLIPUT transforms this step into a single memory access and therefore, incurs a deterministic low latency depending on the decoder configuration. LILLIPUT has low hardware complexity and fits on off-the-shelf FPGAs. As most existing quantum systems~\cite{googlesupremacy,ibmqfpgacontrol} use FPGAs for delivery of control instructions to qubits and implementation of readout interface, it can be seamlessly integrated on these systems. Overall, LILLIPUT is accurate, fast, and lightweight which makes it an ideal candidate for decoding in near-term QEC.

LILLIPUT incurs high memory overhead to store the LUTs and requires a memory external to the FPGAs for some decoder configurations. To address this challenge, we propose \textit{Compressed Look-Up Tables (CLUTs)}. CLUTs exploit the fact that not all entries of a LUT are accessed with equal probability. This behavior arises from the nature of surface codes, where errors on data qubits only flip adjacent parity qubits. Infrequent errors flip a few parity qubits. Alternately, more errors affect multiple locations that are more likely to flip the parity qubits back and forth and result in few bit flips overall. Consequently, the Hamming weight (number of ones) of the memory address accessed is typically low. CLUTs leverage this insight and only store entries corresponding to addresses of low Hamming weights that are most likely to be accessed. CLUTs determine the cut-off Hamming weight such that LILLIPUT encounters a decoder failure due to missing LUT entries with probability equal to or lower than the logical error rate and thus, does not impact the overall accuracy. 

The reconfigurability of LILLIPUT allows us to perform a design space exploration across various decoder configurations and understand the trade-off between the accuracy and complexity of a decoder. Our studies with small surface codes show that LILLIPUT achieves high accuracy. The decoding latency (time from the arrival of syndrome to error assignment) ranges between 29-42 ns for the decoder configurations studied in this paper. LILLIPUT requires up-to 7\% logic utilization on off-the-shelf FPGAs~\cite{intelcyclone,intelarria,intelstratix} and thus, is extremely lightweight. Lastly, CLUTs reduce the memory requirement of LILLIPUT by up-to 107x (from 148 MB to 1.3 MB) without sacrificing decoding accuracy.

\newpage

Overall, this paper makes the following contributions: 

\vspace{0.05 in}
\begin{enumerate}[leftmargin=0cm,itemindent=.5cm,labelwidth=\itemindent,labelsep=0cm,align=left, itemsep=0.3 cm, listparindent=0.5cm]
    \vspace{0.2 cm}
    \item We propose {\em LILLIPUT}, a lightweight low latency lookup table decoder for small surface codes (requires <7\% logic elements on FPGAs). It offers high accuracy as it can tolerate errors on both data qubits as well as in the syndrome extraction circuit; and requires a decoding latency within 42 ns. 
    
    \item To the best of our knowledge, LILLIPUT is the first fully reconfigurable system-level decoding solution that can be seamlessly integrated with existing quantum systems.
    
    \item We propose \textit{Compressed Look-Up Tables (CLUTs)} to reduce the memory overhead of LILLIPUT. By only storing data for the most probable error events, CLUTs reduce the memory required by up-to 107x, without sacrificing the accuracy. 
    
\end{enumerate}

%% file: sections/section2backgroundandmotivation.tex
\section{Background and Motivation}

\subsection{Qubits and Quantum Error Correction}
A quantum bit, or qubit, is the fundamental unit of information in a quantum computer. The state of a qubit can be represented as a superposition of its basis states $\ket{0}$ and $\ket{1}$, with complex valued amplitudes for these states~\cite{nielsen2002quantum}. 
Unfortunately, qubit devices retain information for only a short span of time (about few microseconds) and quantum operations have very high error rates (about 1\%)~\cite{sycamoredatasheet}. These factors limit us from executing most quantum applications on existing hardware. To run programs without encountering errors, quantum information must be protected by using Quantum Error Correction (QEC)~\cite{shor1996fault,aharonov1999fault,aliferis2005quantum,gottesman2010introduction,fowler2012surface}. 

QEC codes encode a logical qubit using a set of data and parity qubits. The data qubits collectively store the quantum information, whereas 
the parity qubits periodically extract information about errors on the data qubits by executing a syndrome extraction circuit. Measuring the parity qubits every QEC cycle allows the QEC code to project any errors on data qubits into a discrete set of Pauli errors. The bit-flip (X) error swaps the probability amplitudes of the basis states, whereas the phase-flip (Z) error introduces a relative phase of -1 between them.
The Pauli Y error denotes simultaneous X and Z errors. The measurement outcomes of the parity qubits, called \textit{syndrome}, is analyzed by a decoder to detect and correct errors encountered by the data qubits. To perform computations, fault-tolerant quantum computers must perform error correction continuously while running an algorithm.

\subsection{Surface Code}
\begin{figure}[htb]
\centering
\vspace{0.01in}
    \includegraphics[width=\columnwidth]{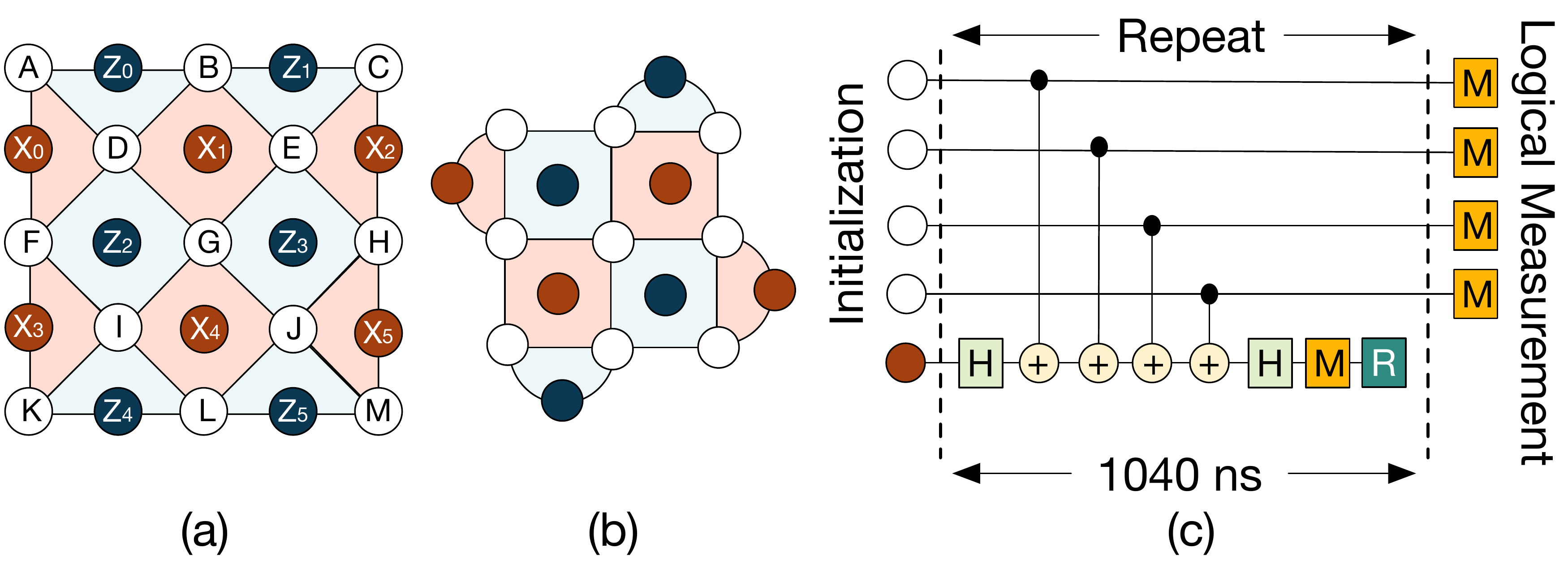}
    \vspace{0.01in}
    \caption{Distance 3 (a) regular and (b) rotated surface code. X errors are detected by the Z stabilizers (in blue) and Z errors are detected by the X stabilizers (in red). (c) Z stabilizer circuit} 
    \label{fig:surfacecode}
\end{figure}

The surface code~\cite{fowler2012towards,kitaev2003fault,dennis2002topological,raussendorf2007fault} is widely considered to be the most promising QEC code as it can tolerate high thresholds~\cite{PhysRevA.89.022321} and requires only nearest-neighbor connectivity. It encodes a logical qubit into a 2-dimensional lattice of alternating data and parity qubits. The size and layout of the lattice depends on the code distance $d$ which determines the code redundancy and the length of the shortest error chain ($\frac{d+1}{2}$) that cannot be corrected. An error on a data qubit 
is detected by the adjacent parity qubits by executing a syndrome extraction or stabilizer circuit, where each parity qubit measures a four qubit operator called a \textit{stabilizer}. X errors are detected by the Z stabilizers, whereas Z errors are detected by the X stabilizers~\cite{PhysRevA.57.127}. For example, Figure~\ref{fig:surfacecode}(a) shows the layout of distance $3$ regular surface code that can correct error chains of length 1. It consists of 13 data qubits (qubits A to M) and 12 parity qubits (qubits $Z_0$ to $Z_6$ and $X_0$ to $X_6$). An X error on data qubit D is detected by the Z stabilizers $Z_0$ and $Z_2$, whereas a Z error on this qubit is detected by the X stabilizers $X_0$ and $X_1$. 

Figure~\ref{fig:surfacecode}(b) shows the layout of distance $3$ \textit{rotated surface code} that is obtained by rotating the lattice of a regular surface code by 45 degrees and removing some of the data and parity qubits. The rotated code requires fewer qubits and gate operations to extract a syndrome. Therefore, the rotated surface code is preferred over regular lattices for near-term experiments, so we focus on rotated codes in this paper.

\subsection{Error Decoding in Real-time} 
A decoder uses the output of stabilizer measurements, the syndrome, to determine a set of corrections that must be applied to the data qubits. By convention, the syndromes generated by the X stabilizers are called X syndromes and are decoded using an X decoder. Similarly, the syndromes generated by the Z stabilizers are called the Z syndromes and decoded using a Z decoder. Decoders must accurately identify errors in real-time to prevent accumulation of errors. The maximum latency that can be tolerated by a decoder is the latency of the syndrome extraction circuit, an example of which is shown in Figure~\ref{fig:surfacecode}(c). 
The circuit has a latency of about 1 $\mu$seconds which is derived existing device technology~\cite{sycamoredatasheet,chen2021exponential}.  
Errors must be corrected within this time and designing accurate and fast decoders is an active area of research. Typically, software implementations are slow~\cite{fowler2017QEC_talk} and hence, more recently decoders have been proposed that uses custom hardware~\cite{das2020scalable} and superconducting devices~\cite{nisqplus,ueno2021qecool} to enable fast decoding. 

\subsection{Motivation: Near-term QEC} 
QEC is essential to realize the potential of quantum computers in practice. As the device quality and system size continues to improve, there is increasing interest in studying QEC codes and real-time decoding. Several demonstrations of repetition codes, Bacon-Shor code have been successful~\cite{cory1998experimental,schindler2011experimental,moussa2011demonstration,zhang2011experimental,reed2012realization,waldherr2014quantum,riste2015detecting,cramer2016repeated,wootton2018repetition,wootton2020benchmarking,chen2021exponential,egan2020fault,luo2020quantum}. While these experiments represent a significant milestone in QEC, unlike surface codes, these codes can only correct either phase-flip or bit-flip errors but not both. Also, QEC experiments so far have mainly relied on offline software-based error decoding. The largest surface code demonstrated so far is distance 2~\cite{chen2021exponential,andersen2020repeated,bell2014experimental,marques2021logical}. With improving device quality, quantum hardware may soon reach the level of 1\% error per data qubit per syndrome cycle, where quantum error correction can function (involving surface codes of distance 3 and beyond). Thus, QEC experiments that demonstrate small surface codes using real-time decoding is a reasonable major milestone for QEC in the next few years. 

\subsection{Challenges in Real-time Decoding}
Real-time decoding is necessary to prevent accumulation of errors on data qubits. If errors are not corrected within a QEC cycle before the arrival of the next syndrome, errors may accumulate resulting in a logical failure. The decoding complexity depends on the error events and software decoders may be too slow for real-time decoding~\cite{fowler2017QEC_talk}. They also incur significant communication overheads in transmitting the syndromes into software running on a CPU. 
This high latency of software decoders limits the use of general purpose computing for online decoding and therefore, almost all QEC experiments that use software decoders have resorted to offline decoding. In the first instance of real-time decoding~\cite{ryan2021realization}, software decoders have been used for color codes~\cite{steane1996error}. However, this study uses trapped-ion systems that can tolerate up to few milli-seconds of decoding latency, about 3 orders of magnitude higher than superconducting systems. The alternative is hardware decoders that are faster and promise real-time decoding. However, they have poor accuracy due to algorithmic and implementation limitations and rely on custom hardware~\cite{das2020scalable} or specialized devices~\cite{nisqplus,ueno2021qecool}. Given the current state of superconducting device technology, it is not even feasible to implement the SFQ decoders~\cite{nisqplus,ueno2021qecool} in the near-term. The number of devices required to fabricate these decoders far exceeds the device densities of existing superconducting technologies.  

\subsection{Goal: Decoding in Real-time for Near-term QEC}
Ideally, we want a low-cost and accurate decoder that can be seamlessly integrated with existing quantum devices while enabling real-time decoding for near-term QEC experiments. To achieve this goal, we propose LILLIPUT-- an accurate Lightweight Low Latency Look-Up Table decoder for small surface codes in this paper.

%% file: sections/section3methodology.tex
\section{Evaluation Methodology}
\label{sec:evaluation}
In this section we describe the benchmarks, experimental setup, and the figure-of-merit used to evaluate our policies before discussing our design. 
\subsection{Surface Code Parameters} 
In this paper, we consider rotated surface codes of distance 3, 4, and 5. The details of the layouts are described in Table~\ref{tab:layoutinformation}. The total number of physical qubits required ranges from 17 to 49. In the near-term, we expect systems with few hundreds of qubits which would be able to fit these layouts.

\begin{table}[htb]
\begin{center}
\begin{small}
\caption{Parameters of Rotated Surface Codes}

\setlength{\tabcolsep}{1.2mm} 
\renewcommand{\arraystretch}{1.2}
\label{tab:layoutinformation}
{\footnotesize
\begin{tabular}{ |c|c|c|c|c|} 
\hline
Code & Data & X-ancilla & Z ancilla & Total Physical \\ 
Distance & Qubits & Qubits & Qubits & Qubits \\
\hline
\hline
3 & 9 & 4 & 4 & 17 \\
\hline
4 & 16 & 8 & 7 & 31 \\
\hline
5 & 25 & 12 & 12 & 49\\
\hline

\end{tabular}}
\end{small}
\end{center}
\end{table}

\subsection{Monte Carlo Simulation Infrastructure}
\begin{figure}[htp]
\centering
    \includegraphics[width=\columnwidth]{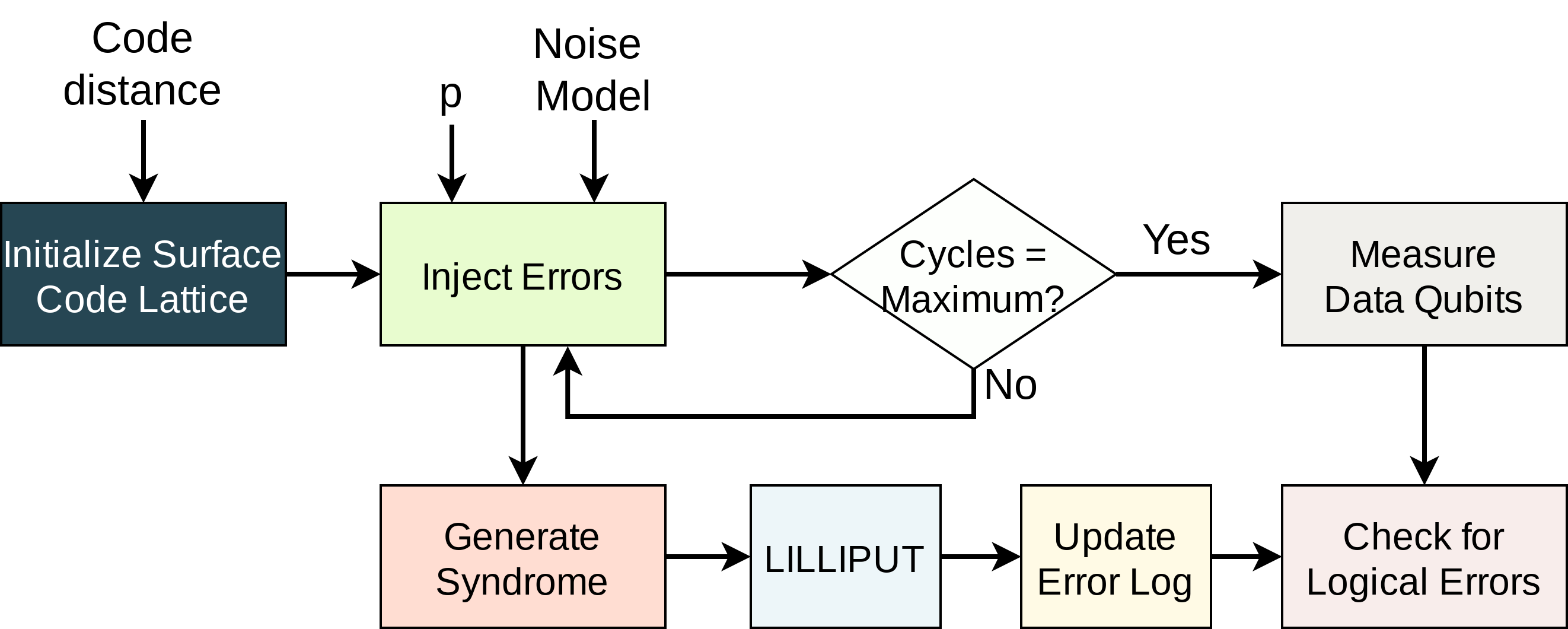}
    \caption{Monte Carlo simulation framework} 
    \label{fig:montecarlosim}
\end{figure}

Figure~\ref{fig:montecarlosim} shows an overview of the Monte Carlo simulator used for our studies. The simulator generates a surface code lattice for a given code distance. Depending upon the noise model and the physical error rate $p$, the simulator injects errors onto the data qubits and measurement errors onto the parity qubits, producing a syndrome every cycle. The X and Z syndromes are then decoded independently and an error log is maintained for both error types. To model a QEC experiment, the simulator repeats the process for multiple cycles and terminates when the maximum number of cycles is reached. The simulator maintains the internal state of the qubits throughout the experiment which is then used to compute the logical measurement outcome and the logical error. We call each such execution as a \textit{trial}. For our evaluations, we use 1 million random trials. The simulator also generates the traces used to verify the proposed LILLIPUT micro-architecture. 

\begin{figure*}[!tp]
\centering
    \includegraphics[width=5 in]{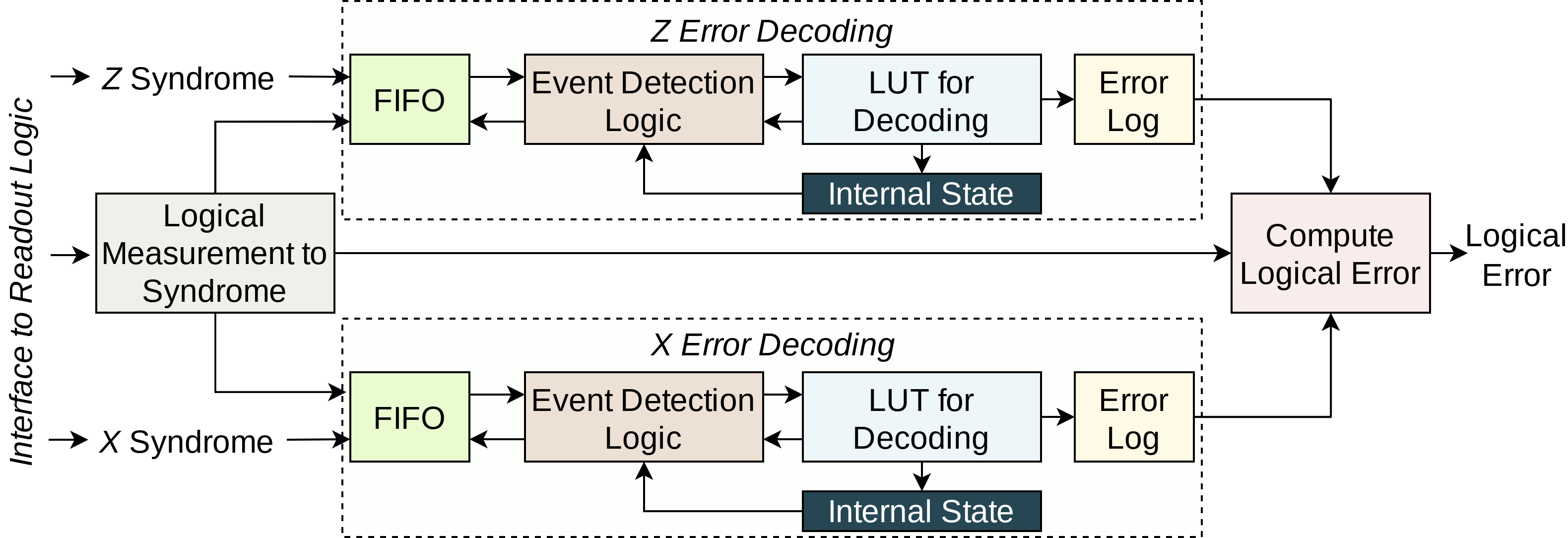}
    \caption{Overview of LILLIPUT} 
    \label{fig:overview}

\end{figure*}

\subsection{Noise Model}
We implement the \textit{phenomenological noise model}~\cite{dennis2002topological}, which inserts errors on both data qubits and on the measurement of parity qubits. This is a standard noise model used in QEC and has been used in several prior works. In this noise model, a data qubit encounters an error with probability $p$ in each cycle. The type of error is chosen uniformly from Pauli X, Y, and Z errors. Additionally, each parity qubit encounters measurement error with probability $p$. For simplicity, we assume the probabilities for data qubit errors to be the same as measurement errors. This assumption is consistent with prior works in QEC. For our studies, we consider physical error rates ranging from $p = 10^{-3}$ to $p = 5\times10^{-2}$. We consider this to be a suitable range of error rates for quantum devices in the near term. Nonetheless, if the device quality improves further, our design can still support those quantum architectures.

\subsection{Target Hardware Platforms}
Our target is to implement the decoder on commercially available FPGAs as existing quantum systems already use FPGAs for control and readout interface logic~\cite{googlecryocontrol,googlesupremacy,ibmqfpgacontrol}. For our studies we use FPGA from the Intel Cyclone 10 LP~\cite{intelcyclone}, Arria V~\cite{intelarria}, and Stratix 10~\cite{intelstratix} family as these are commercially available. 

%% file: sections/section4design.tex
\section{Overview of LILLIPUT}
The classical counterpart of QEC comprises of three key steps-- \textbf{(1)} error detection from the stabilizer measurements, \textbf{(2)} identification of errors and error assignment to data qubits every cycle, and \textbf{(3)} computation of logical error. The micro-architecture of LILLIPUT shown in Figure~\ref{fig:overview} accomplishes these steps. It communicates with the readout interface, translates the stabilizer measurement outcomes into error detection events, identifies errors, and computes the logical error. Note that X and Z errors are decoded independently. In the next subsections, we describe the implementation of the design.

\subsection{Detection of Errors from Stabilizer Measurements}
\begin{figure}[htb]
\centering
    \includegraphics[width=\columnwidth]{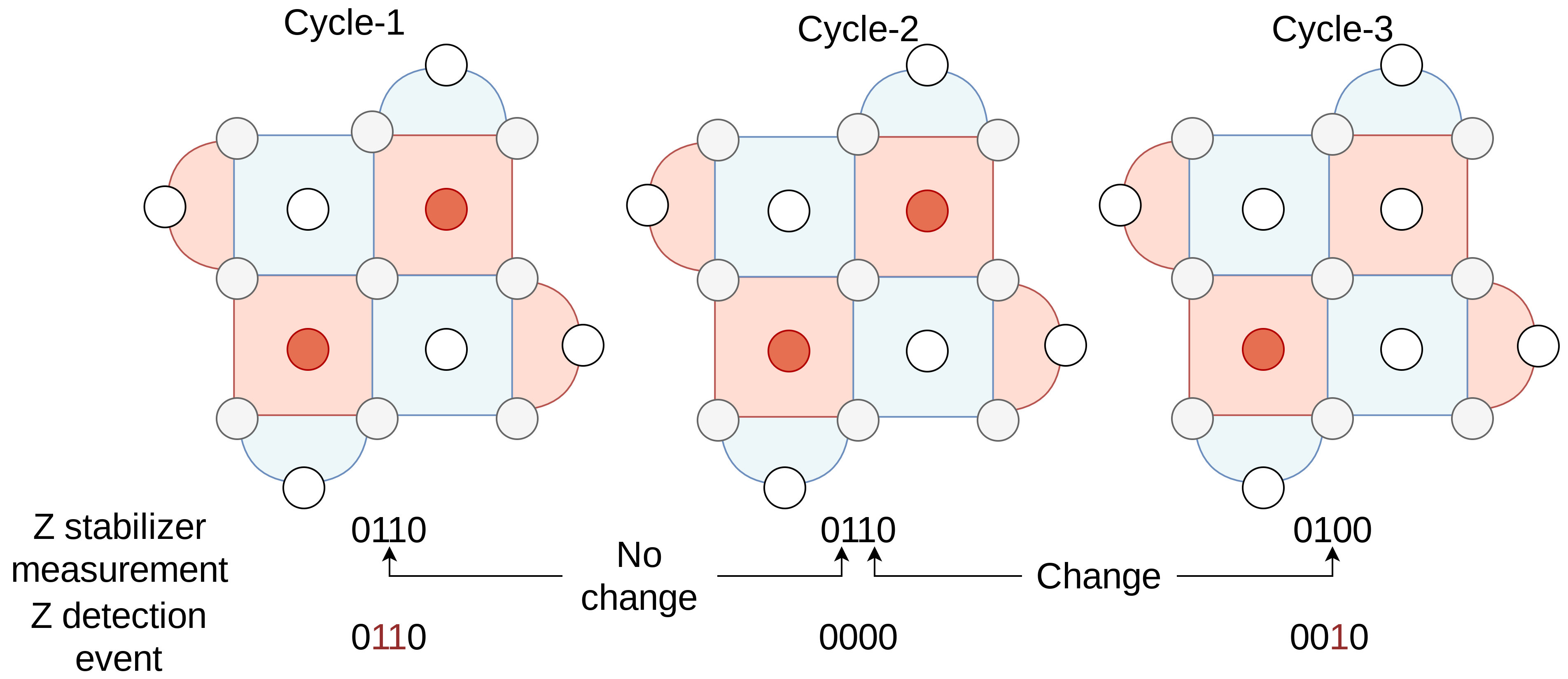}
    \caption{Steps involved in translating the stabilizer measurement outcomes (red denotes an error is identified) to error detection events in each QEC cycle. The bitstream for stabilizer measurements is specified from left to right (convention used in this paper) on the surface code lattice.} 
    \label{fig:detectionevent}
\end{figure}

In QEC, syndromes are generated every cycle by measuring the stabilizers. LILLIPUT generates \textit{error detection events} by comparing the stabilizer measurement outcomes from two consecutive QEC cycles. Any change in the measurement outcome of the stabilizers between two cycles indicates an error. Alternately, no event is detected if the stabilizer measurement outcomes remain the same. For example, Figure~\ref{fig:detectionevent} shows Z stabilizers measurements outcomes of a distance 3 surface code, where an error event is detected in cycles 1 and 3, whereas no error event is detected in cycle 2. This step is accomplished by the Event Detection Logic block shown in Figure~\ref{fig:overview}. Detection events enables us to track errors in a given cycle and are used to identify the optimal correction. 

\subsection{Error Identification as a Matching Problem}
Error identification is the step in which a decoder assigns errors to each data qubit which is tracked throughout the QEC cycles. To perform this step, the detection events are represented on a graph, often termed as the \textit{decoding graph}, where the nodes and edges represent the parity and data qubits respectively. The minimum weight perfect matching algorithm~\cite{fowler2012MWPM} uses the decoding graph and matches each detection node with another or to the surface code boundary such that the total weight of the matched edges is minimal. MWPM is considered to be the most accurate algorithm. \footnote{A more recent approach, the Union-Find algorithm~\cite{delfosse2017linear,delfosse2017almost}, uses a different approach to matching to generate the error assignments. It is faster than MWPM but has lower accuracy~\cite{huang2020WUF}.} In LILLIPUT, this step is accomplished using Look-Up Tables (LUTs), as shown in Figure~\ref{fig:overview}, and the details are described in Section~\ref{sec:decoding}. 

\subsection{Handling Data and Measurement Errors}
For greater accuracy, a decoder must handle both errors on the data qubits as well as errors in the syndrome extraction circuit. Overall, there are four key sources of errors: (a)~errors on data qubits (b)~errors in the gate operations during syndrome extraction (c)~measurement errors on parity qubits during stabilizer measurements, and (d)~measurement errors on data qubits during logical measurement. Next, we discuss how LILLIPUT handles each of these errors, where we use ``space'' and ``time'' directions to describe when detection events are generated in the same round or two consecutive cycles, respectively.

\begin{figure}[htp]
\centering
    \includegraphics[width=\columnwidth]{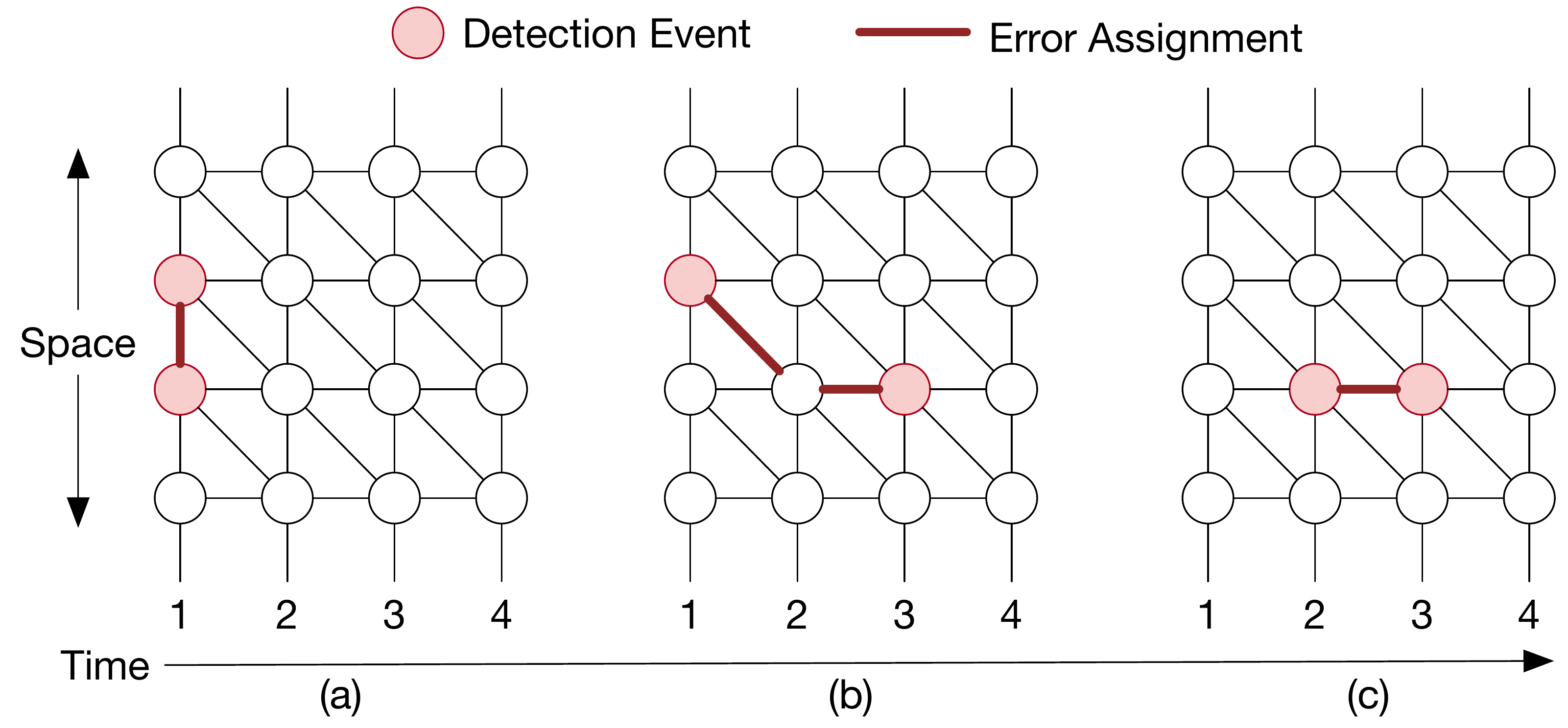}
    \caption{A distance 4 repetition code decoding graph, where nodes and edges denote parity and data qubits respectively. Note that this is for illustration only and our evaluations in this paper are based on surface codes. Examples of (a) space-like (b) space-time like and (c) time-like detection events.} 
    \label{fig:decodinggraph}
\end{figure}

\vspace{0.05 in}
\noindent \textbf{(a)} An error on a data qubit is detected by its neighboring parity qubits, producing a \textbf{\textit{space-like}} detection event on them in the same cycle. For example, Figure~\ref{fig:decodinggraph}(a) shows a space-like error detection event in cycle 1. 

\vspace{0.05 in}
\noindent \textbf{(b)} Two-qubit gate errors during syndrome extraction are detected across consecutive cycles. They produce \textbf{\textit{space-time like}} detection events, as shown in cycles 1 and 3 in Figure~\ref{fig:decodinggraph}(b). 

\vspace{0.05 in}
\noindent \textbf{(c)} Measurement errors on parity qubits exhibit temporal behavior and translates into \textbf{\textit{time-like}} detection events in two consecutive cycles, as shown in cycles 2 and 3 of Figure~\ref{fig:decodinggraph}(b). 

\vspace{0.05 in}
\noindent \textbf{(d)} Measurement errors on the data qubits are handled in the last boundary cycle by generating an extra syndrome of either X or Z type and is discussed in detail in the next subsection.

\vspace{0.05 in}
\noindent To tackle the first three sources of errors, LILLIPUT processes more than a single cycle of syndrome and performs matching on a 3-dimensional decoding graph spanning space and time (multiple cycles). For errors in logical measurement, LILLIPUT uses an extra syndrome as discussed in Section~\ref{sec:boundary}. 

LILLIPUT assumes the readout logic writes the measurement outcomes on a buffer which gets overwritten every cycle. This interface works on a much slower clock (e.g. with 1 $\mu s$ latency) that depends on the duration of the syndrome extraction circuit. LILLIPUT polls for new syndromes and tracks the most recent history in a FIFO, as shown in Figure~\ref{fig:overview}. The FIFO must store the history of the last $m$ cycles, where $m$ is the number of syndrome rounds that are simultaneously decoded. Since LILLIPUT decodes in real-time, storing only the last $m$ rounds is sufficient. We discuss the impact of number of measurement rounds on the logical error rate in Section~\ref{sec:evaluations}. 

\subsection{Handling Boundary Cycles}
\label{sec:boundary}
A QEC experiment has two time boundary cycles- one at the beginning and the other at the end. \footnote{In QEC, cycles are also called \textit{rounds}. We use the term cycles in this paper to avoid confusion with the number of syndrome rounds used for decoding.} For the beginning time boundary cycle, the detection event is computed by comparing the first round of stabilizer measurements and the qubit initialization data. In the last time boundary cycle, the data qubits are measured. To tolerate measurement errors in data qubits, LILLIPUT translates the measurement outcomes of the data qubits into a detection event of either X or Z type by comparing with the stabilizer measurement data from the second-last cycle. Since a logical measurement in surface codes is implemented by performing a transversal measurement (measuring all the data qubits in either X or Z basis), the measurement basis of the data qubits determines the type of the stabilizer that can be constructed from the measurement outcomes. By default, LILLIPUT translates the data qubit measurement outcomes into a Z syndrome because we assume the data qubits are measured in the Z basis, as shown in Figure~\ref{fig:overview}. 

%% file: sections/section5lut.tex
\section{Error Assignments in LILLIPUT}
\label{sec:decoding}
LILLIPUT determines the most optimal error assignment by using Look-Up Tables (LUTs).
Every cycle, the history of detection events is used to index an LUT and the LUT entry assigns errors to the data qubits in the oldest cycle. The length of the history depends on the number of syndrome rounds used for decoding. LILLIPUT also maintains an \textit{internal state} which is updated every cycle. 
The details of using LUTs in LILLIPUT and tracking the internal state is described next.

\subsection{Streaming Mode of Operation}
\begin{figure}[htp]
\centering
    \includegraphics[width=0.9\columnwidth]{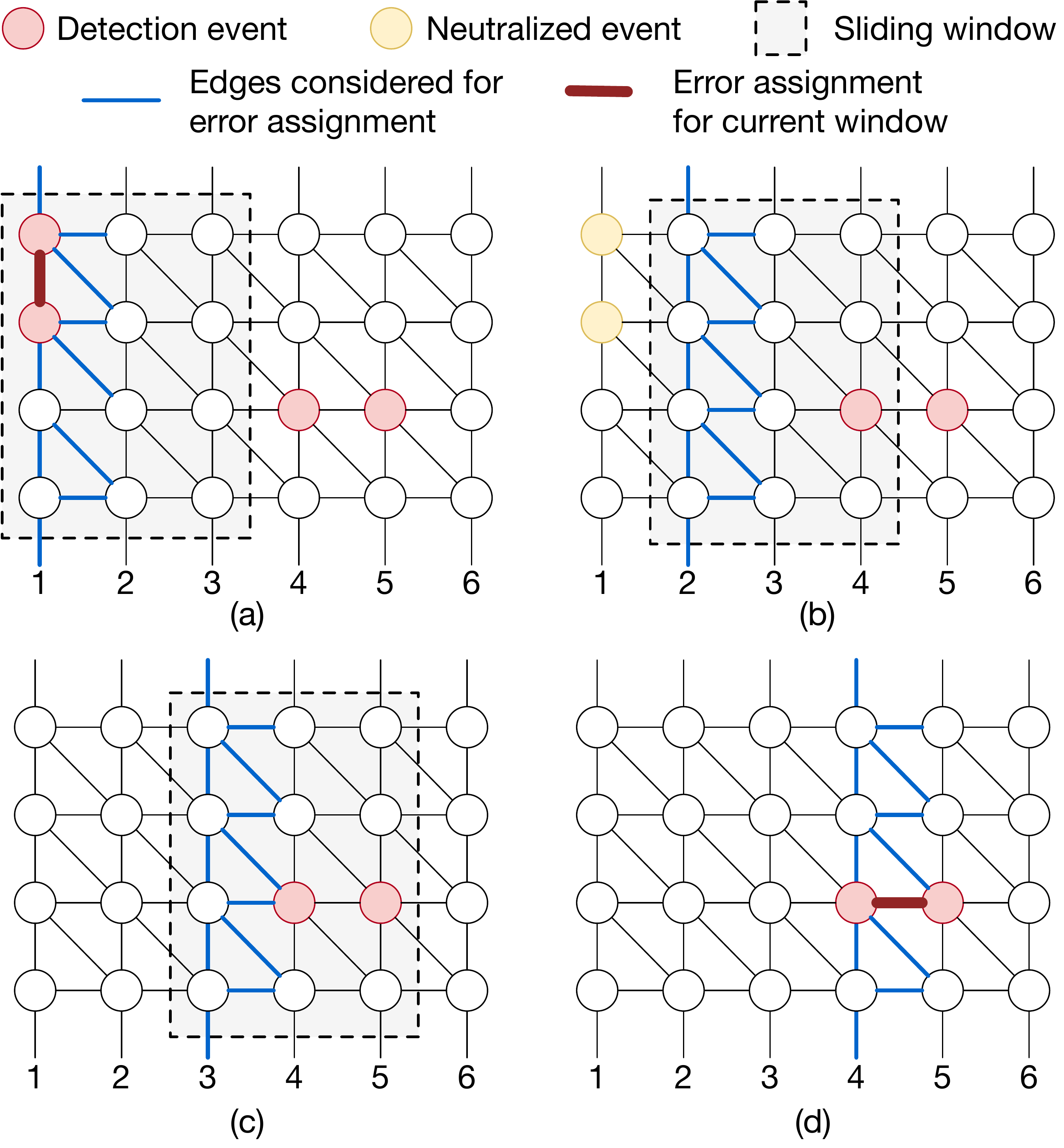}
    \caption{(a) The sliding window uses detection events from cycles 1, 2, and 3 and assigns error to the oldest layer i.e. cycle 1. Note that all detection events in the  window are considered for matching but only the edges touching the oldest cycle are considered during error assignments, shown in blue. (b-d) The sliding window streams forward one cycle at a time.} 
    \label{fig:slidingwindow}
\end{figure}

LILLIPUT operates in \textit{streaming} mode and maintains an \textit{error log} for each data qubit. The log is updated every cycle as errors are identified. LILLIPUT considers a fixed number of cycles at a time which is equal to the number of syndrome rounds used in decoding. We call this a \textit{sliding window} because it slides forward as an experiment proceeds. For example, Figure~\ref{fig:slidingwindow} shows the sliding window for four consecutive decoding steps. LILLIPUT uses the detection events in the sliding window and the internal state to determine the LUT address every cycle. 

\subsection{Error Assignment}
LILLIPUT determines the best correction by using all the detection events in each sliding window and assigns errors only to the oldest layer. For example, Figure~\ref{fig:slidingwindow} shows the subset of edges that the decoder uses to make the error assignment for the current sliding window. The optimal error assignment is stored in the LUTs and targets to \textit{neutralize} the effect of all of the errors in the oldest cycle and decoding completes once the sliding window covers all the syndromes. The error assignments are made only to the oldest layer in each sliding window to prevent any \textit{premature matching}. Premature matching may result in sub-optimal performance if syndromes in the future cycles can change the assignment. Premature matching may also result if we use disjoint windows where the window proceeds forward by its full length. In that case, errors that span over two non-overlapping windows can cause inaccurate error assignments. As LILLIPUT uses a sliding window, it does not encounter this problem and therefore, does not suffer from sub-optimal performance. For the time boundary cycles, LILLIPUT constructs a full sliding window by padding zeros. Figure~\ref{fig:detectioneventaddition}(c) shows a time boundary cycle where additional zeros are padded. Since the LUT is programmed such that it assigns the most optimal error for a sliding window, it is guaranteed to never assign errors to the zero-padded regions.

\subsection{Syndrome Updates and Tracking the Internal State}
\begin{figure}[htp]
\centering
    \includegraphics[width=\columnwidth]{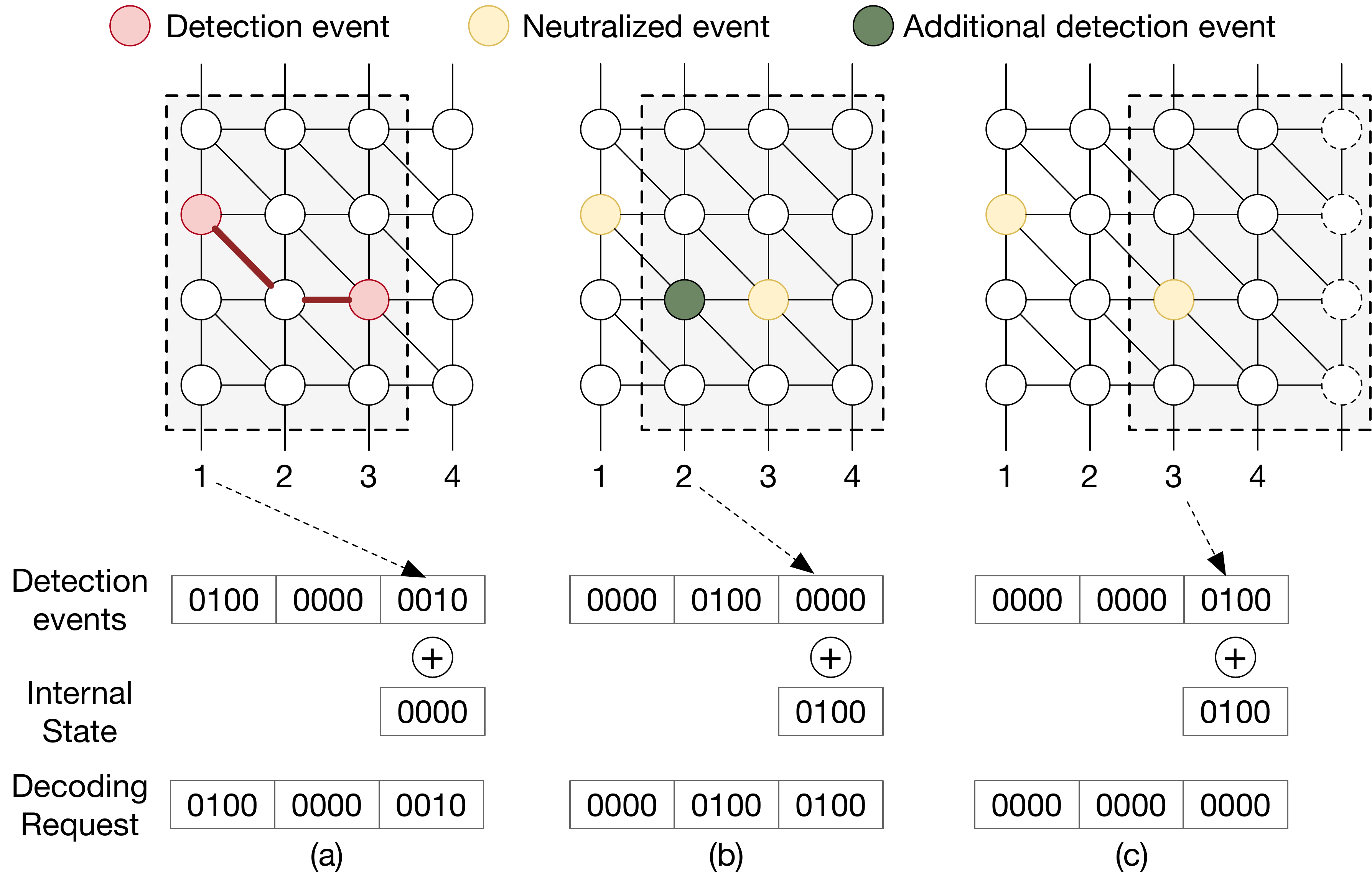}
    \caption{(a) LILLIPUT finds the optimal matching in current window but adds a detection event in cycle 2 which is tracked in the internal state register. (b-c) The additional event is taken into account from the register. In the boundary cycle, zeros are padded. By convention, the detection events of the oldest cycles occupy the least significant bits. The bits for each cycle run bottom to top from most significant position to the least.} 
    \label{fig:detectioneventaddition}
\end{figure}

LILLIPUT assigns errors only to the oldest layer. However, as it performs matching across all the events in a sliding window, it can neutralize some errors in the second oldest cycle if the detection event has a time-like or space-time like behavior. For example, Figure~\ref{fig:slidingwindow}(d) shows a time-like detection event that requires neutralizing events in both cycles 4 (oldest cycle) and 5 (second oldest cycle). Consequently, matching within this window removes the detection event from the second oldest layer. Similarly, detection events may be added as well. For example, Figure~\ref{fig:detectioneventaddition}(a) shows an example of decoding steps where detection events are added. To accommodate these scenarios, LILLIPUT maintains a record of the detection events added or removed in an internal state register. To obtain the LUT address to be accessed in a cycle, LILLIPUT concatenates the detection events in the current window and modifies the oldest detection event using the most recent internal state, as shown in Figure~\ref{fig:detectioneventaddition}. Each LUT entry provides the error assignment for the oldest layer as well as the detection events that are added or removed in the current cycle (1 bit per parity qubit). This value is used to update the internal state register every cycle.  

\subsection{Programming the LUT}
\begin{figure}[htp]
\centering
    \includegraphics[width=\columnwidth]{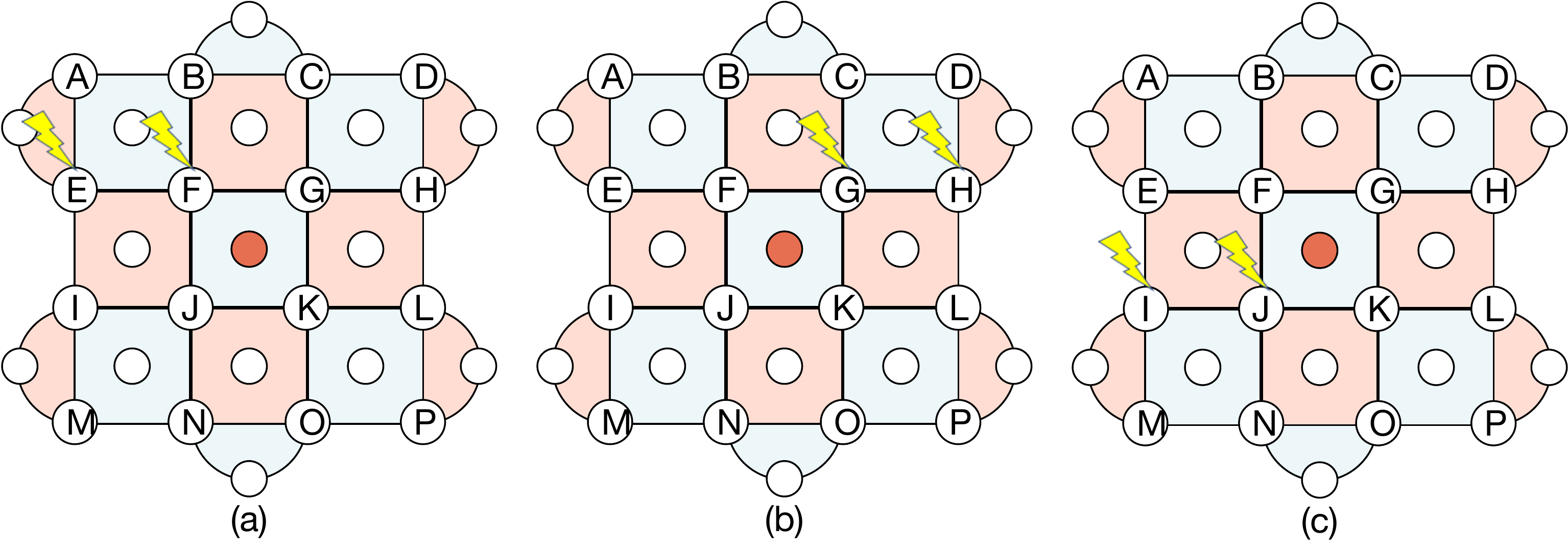}
    \caption{Valid possible error assignments for the detection event at the center of the distance 4 surface code lattice.} 
    \label{fig:multipleassignment}
\end{figure}

LILLIPUT programs the LUT entries offline by generating all possible detection events for a given code distance and size of the sliding window. We use the software Minimum Weight Perfect Matching (MWPM) decoder~\cite{fowler2013minimum}. For some events, the MWPM decoder may provide more than one possible error assignment. In other words, a single detection event may have multiple matching possibilities on the decoding graph that result in minimal weight. For example, Figure~\ref{fig:multipleassignment} shows possible error assignments on a distance 4 surface code lattice for the same detection event. Here, the Z stabilizer at the center of the lattice indicates an error. The MWPM decoder can assign X errors to either data qubits (a) E and F or (b) G and H or (c) I and J.  For high accuracy, we account for the error model of the quantum hardware (we consider Google Sycamore~\cite{chen2021exponential}) to select the most probable error assignment. LILLIPUT can also accommodate variability in device error rates~\cite{tannu2018case,noiseadaptive} on existing systems by re-programming the LUTs. As device characteristics remain stable over short periods of time and mainly exhibit variability over extended periods (weeks or months)~\cite{dasgupta2021stability}, the LUTs need not be re-programmed too frequently. The reconfigurability allows LILLIPUT to be adapted to other QEC codes, decoding algorithms, and quantum systems. 

\subsection{Determination of Logical Error}
Early demonstrations of QEC will perform a memory experiment, simply showing that error correction preserves the initial quantum state.  The probability of logical error is determined by comparing the logical measurement outcome with the expected outcome for the state prepared.  This logical measurement is computed using the error log. Which error log (X or Z) is used depends on the measurement basis of the logical measurement. The logical measurement bit is computed by using a reduction XOR operation on the bitwise XOR results of the error log and the logical measurement outcome. 

%% file: sections/section6results.tex
\section{Final Evaluations}
\label{sec:evaluations}
In this section, we discuss the accuracy, latency, and hardware complexity of our decoder. We consider a wide range of decoder configurations for our evaluations to understand the trade-off between the complexity and accuracy of a decoder. 

\subsection{Results for Accuracy} 
\begin{figure}[htb]
\centering
    \includegraphics[width=\columnwidth]{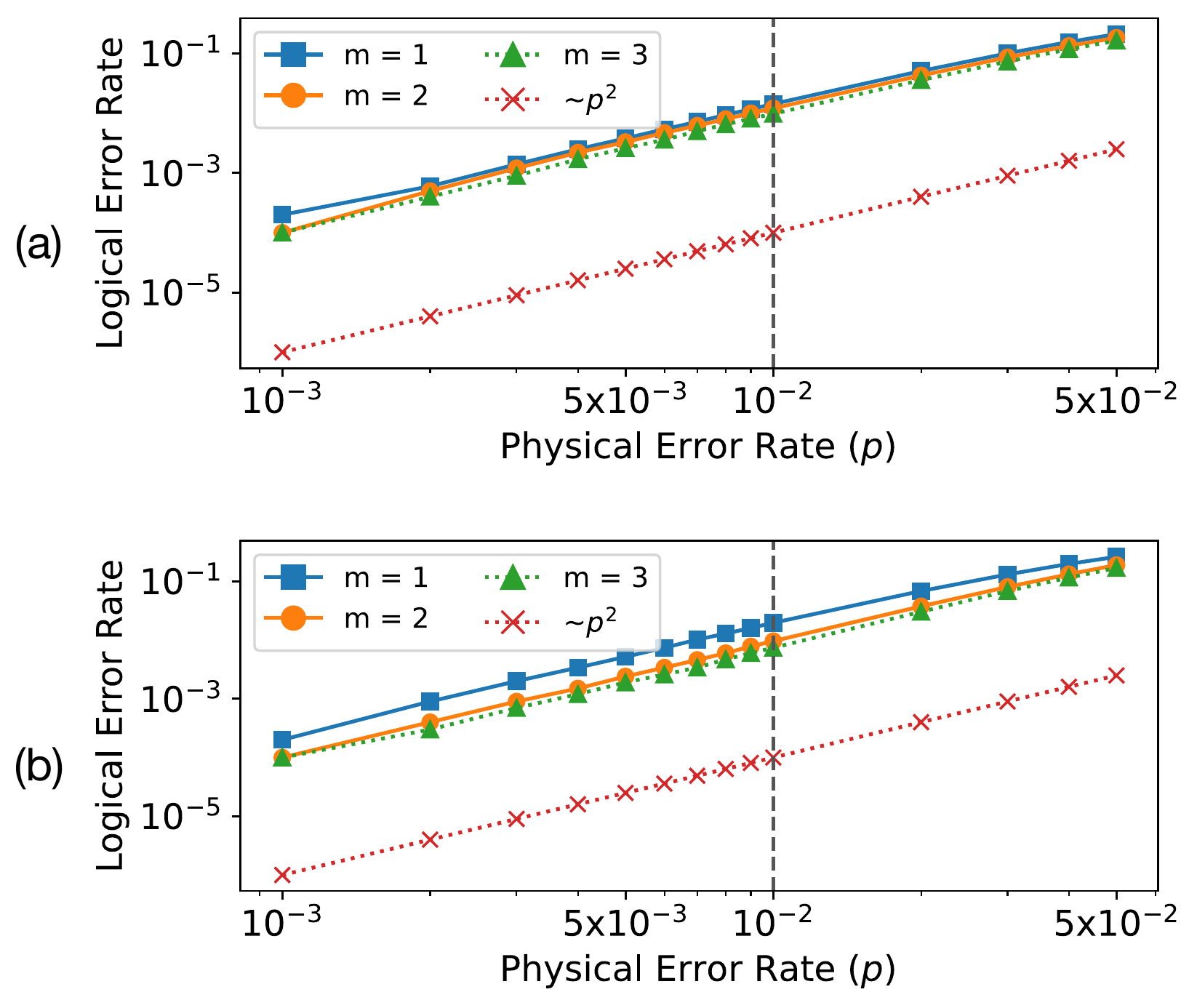}
    \caption{Accuracy of distance (a) 3 and (b) 4 surface codes for different number of syndrome measurement rounds (m)} 
    \label{fig:distance4accuracy}
\end{figure}

Figure~\ref{fig:distance4accuracy} shows the logical error rate (LER) for distances ($d$) 3 and 4 respectively. By default, we assume 5 cycles in the experiments. The LER scales O($p^2$) which is expected because distances 3 and 4 can correct at least one error, but they sometimes fail with two errors; quadratic scaling results from errors being independent in the model used for simulation. We also study the impact of the number of syndrome rounds on decoding accuracy. While $m = (d-1)$ rounds are needed to detect as many errors as the code is capable of correcting~\cite{PhysRevA.89.022321}, we observe that LILLIPUT performance saturates at $m=2$ syndrome rounds for $d=4$. The LER reduces by 1.23x and 1.21x on average for distance 3 by going from 1 to 2 and 2 to 3 rounds respectively. Similarly, the LER reduces by 1.99x and 1.24x on average for distance 4 by going from 1 to 2 and 2 to 3 rounds respectively.

\begin{figure}[htb]
\centering
    \includegraphics[width=0.95\columnwidth]{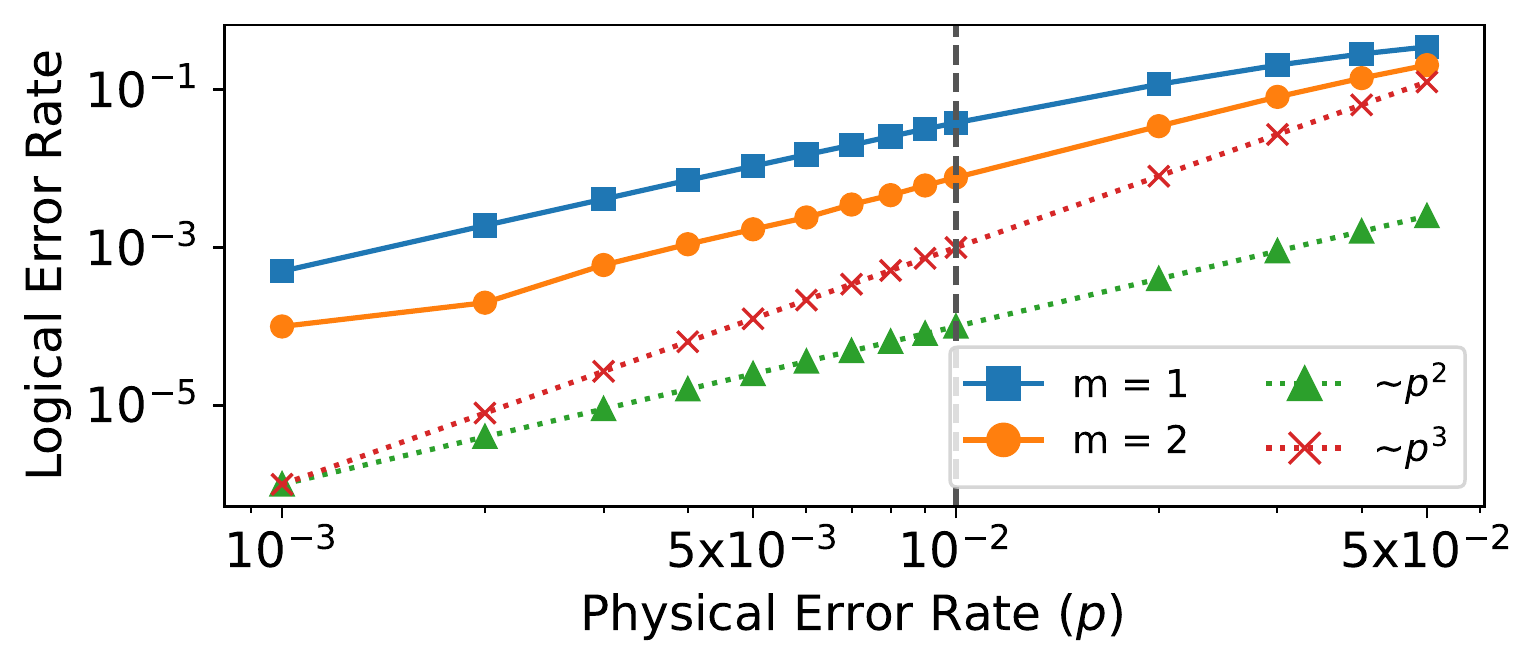}
    \caption{Accuracy of distance 5 surface codes for different number of syndrome measurement rounds (m)} 
    \label{fig:distance5accuracy}
\end{figure}

\begin{figure}[!htb]
\centering
    \includegraphics[width=0.95\columnwidth]{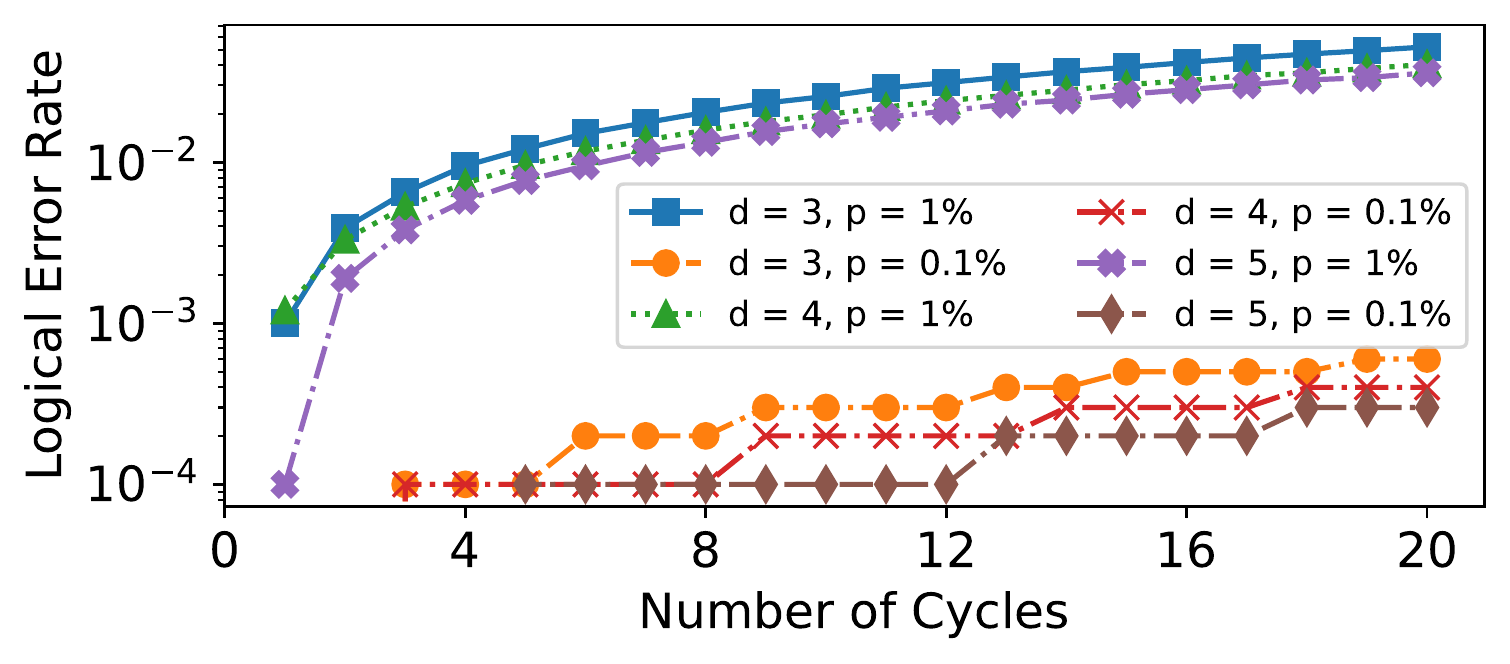}
    \caption{Impact of number of cycles on LER} 
    \label{fig:cyclesvsler}
\end{figure}

Figure~\ref{fig:distance5accuracy} shows the logical error rate for distance 5 surface codes using 1 and 2 rounds of syndrome measurements. We restrict the number of rounds to 2 to keep the LUT sizes tractable for our simulations. We observe that the LER scales O($p^2$), which is expected as using just $m=2$ rounds will lead to some configurations of two errors not being accurately corrected by the decoder.  We also study the impact of number of cycles on the LER which is shown in Figure~\ref{fig:cyclesvsler}. This allows us to budget cycles in a QEC experiment based on the device error rates by accounting for the decoder performance.

\subsection{Results for Hardware Complexity}
The key component of LILLIPUT are the LUTs for both error types (X and Z). The size of each LUT depends on \textit{decoder configuration}. We denote a decoder configuration as $[d,m]$, where $d$ and $m$ are the code distance and number of syndrome rounds considered for decoding respectively. The address size is the length of the detection event which is equal to the syndrome length multiplied by the number of syndrome rounds.
The length of each LUT entry is equal to the sum of the number of data (for error assignment) and parity qubits (for internal state). Table~\ref{tab:lutsize} shows the size of LUTs for different decoder configurations. Asymmetry in the code leads to X and Z syndrome lengths being different for distance 4, so the LUT sizes are different for X and Z syndromes. We use Cyclone 10 LP FPGAs for the decoder configurations $[d=3,m=2]$, $[d=3,m=3]$, and $[d=4,m=2]$ as it only supports up to 486 KB of embedded memory~\cite{intelcyclone}. To support configurations with larger LUTs such as $[d=4,m=3]$, and $[d=5,m=2]$, we use Arria V FPGAs where the LUTs are accessed from an external SRAM (using QDR II) or SDRAM (using DDR2)~\cite{intelarria}. Alternately, the LUTs can be compressed (discussed in Section~\ref{sec:clut}) and does not require external memory~\cite{intelstratix}. 

\begin{table}[htb]
\begin{center}
\begin{small}
\caption{LUTs for different decoder configurations}

\setlength{\tabcolsep}{1.2mm} 
\renewcommand{\arraystretch}{1.2}
\label{tab:lutsize}
{\footnotesize
\begin{tabular}{ |c|c|c|c|c|c|} 
\hline
Decoder & Address & Entry & LUT & Total & Memory \\ 
Configuration & Size & Size & Size & Memory & Type \\
\hline
\hline
$[d=3, m=2]$ & 8 & 13 & 416 B & 832 B & Embedded   \\
\hline
$[d=3, m=3]$ & 12 & 13 & 6.5 KB & 13 KB & Embedded \\
\hline
$[d=4, m=2]$ & 14/ 16 & 23/ 24 & 46/ 192 KB & 238 KB & Embedded \\
\hline 
$[d=4, m=3]$ & 21/ 24 & 23/ 24 & 5.75/ 48 MB & 53.75 MB & External \\
\hline
$[d=5, m=2]$ & 24 & 37 & 74 MB & 148 MB & External \\
\hline

\end{tabular}}
\end{small}
\end{center}
\end{table}

The FPGA utilization is mentioned in Table~\ref{tab:logicutilization}. The logic utilization for LILLIPUT is less than 7\%, making it extremely lightweight and leaving enough room for other circuits such as readout interface logic and logic for delivering control instructions to the qubits. The LUTs consume up to 40\% of the memory bits for designs that use embedded memory. We use different Cyclone devices for distance 3 and $[d=4,m=2]$ for higher performance. Configuring the distance-3 design on the latter FPGA reduces the maximum frequency to 245 MHz.

\begin{table}[htb]
\begin{center}
\begin{small}
\caption{Post synthesis results for different configurations}

\setlength{\tabcolsep}{1.2mm} 
\renewcommand{\arraystretch}{1.2}
\label{tab:logicutilization}
{\footnotesize
\begin{tabular}{ |c|c|c|c|c|c|} 
\hline
Decoder & FPGA & Total LEs/ & Total & \multicolumn{2}{c|}{Utilization} \\ 
\cline{5-6}
Configuration & Family & ALMs & Registers & Area & Memory \\
\hline
\hline
$[d=3, m=2]$ & Cyclone 10 & 353 & 209 & 6\% & 1\% \\
\hline
$[d=3, m=3]$ & Cyclone 10 & 418 & 239 & 7\% & 21\%\\

\hline
$[d=4, m=2]$ & Cyclone 10 & 557 & 340 & < 1\% & 40\%\\
\hline
$[d=4, m=3]$ & Arria-V & 217 & 409 & < 1\% & --\\

\hline
$[d=5, m=2]$ & Arria-V & 246 & 486 & < 1\% & --\\
\hline
\end{tabular}}
\end{small}
\end{center}
\end{table}

\subsection{Results for Latency}
LILLIPUT requires only a single memory access to the LUT and incurs fixed latency, irrespective of the error detection event. LILLIPUT is fully pipelined and requires 7 cycles to determine the correction after a cycle of stabilizer measurements are received. The maximum clock-frequency values depend on the decoder configuration and are listed in Table~\ref{tab:decodinglatency}. To compute the latency of the decoder configurations that rely on external memory access, we account for the most conservative off-chip memory access time (slowest clock frequency and maximum number of cycles) and add it to the latency of the rest of the logic. The decoding latency is up-to 24x lower than the target latency of $1 \mu$seconds in the worst-case (considering the largest decoder configuration studied in this paper). LILLIPUT also meets the 400 ns target latency considered in few prior works~\cite{nisqplus,ueno2021qecool,das2020scalable} and thus, will remain useful even if the duration of syndrome extraction decreases in future with improving device quality.

\begin{table}[htb]
\begin{center}
\begin{small}
\caption{Maximum Frequency (in MHz) and Latency (in ns)}

\setlength{\tabcolsep}{1.2mm} 
\renewcommand{\arraystretch}{1.2}
\label{tab:decodinglatency}
{\footnotesize
\begin{tabular}{ |c|c|c|c|c|c|} 
\hline
\multirow{2}{*}{Metric} & \multicolumn{5}{c|}{Decoder Configurations}\\
\cline{2-6}
                        & [d=3,m=2] & [d=3,m=3]  & [d=4,m=2]  & [d=4,m=3]  & [d=5,m=2]  \\ 
\hline
\hline
Frequency & 250 & 240.7 & 209.8 & 244.4 & 232.9 \\
\hline
Latency & 28 & 29.1 & 33.4 & 40.8 & 42 \\
\hline 
\end{tabular}}
\end{small}
\end{center}
\end{table}

The low logic utilization and re-usability of the LUTs allows LILLIPUT to support QEC experiments spanning more than one logical qubit. As there is sufficient slack between the decoding latency and the target latency for all the decoder configurations studied in this paper, the real-time decoding capability will not be impacted.

%% file: sections/section7clut.tex
\section{A Case for Compressed LUTs}
\label{sec:clut}
The size of a lookup table increases exponentially in the code distance because there is an entry for every possible syndrome bitstring.  This limits the scalability of LILLIPUT, and some of the decoder configurations studied in this paper rely on memory external to the FPGAs. Moreover, it is infeasible to scale the LUT to higher configurations, $[d=5,m=4]$ for example. To implement these configurations without external memory or scale LILLIPUT to other higher configurations, we propose the use of Compressed LUTs (CLUTs).

\subsection{Insight: Not all Error Events are equally likely}
Each LUT entry corresponds to an error event. However, not all error events are equally likely, and Compressed LUTs exploit this using the following insights:

\vspace{0.05 in}
\begin{enumerate}[leftmargin=0cm,itemindent=.5cm,labelwidth=\itemindent,labelsep=0cm,align=left, itemsep=0.1 cm, listparindent=0.4cm]

    \item On surface codes, errors are detected by neighboring parity qubits. When error rates are low, fewer bit flips are observed in the detected events. Alternately, higher error-rates cause errors in multiple locations that result in longer chain of errors. However, it results in only few bit flips overall as the parity qubits inside the chain are likely to be zeros (can be thought of as flipped twice). Thus, the average Hamming weight of the accessed memory addresses is low. For example, the $[d=3,m=2]$ configuration requires LUTs with 8-bit address ranging from 0x00 to 0xFF. Figure~\ref{fig:hammingweightdist3} shows the probability distribution of the Hamming weight of addresses accessed over 1 million trials for different error rates and both X and Z LUTs of this decoder configuration.  We observe that not all addresses are accessed with equal probability and some are seldom or never accessed (address 0xFF for example).
    \textit{Therefore, the memory required for the LUTs can be reduced by removing entries that are unlikely to be accessed. }
    
    \begin{figure}[htp]
    \centering
    \subfloat[\centering]{{\includegraphics[width=4.2cm]{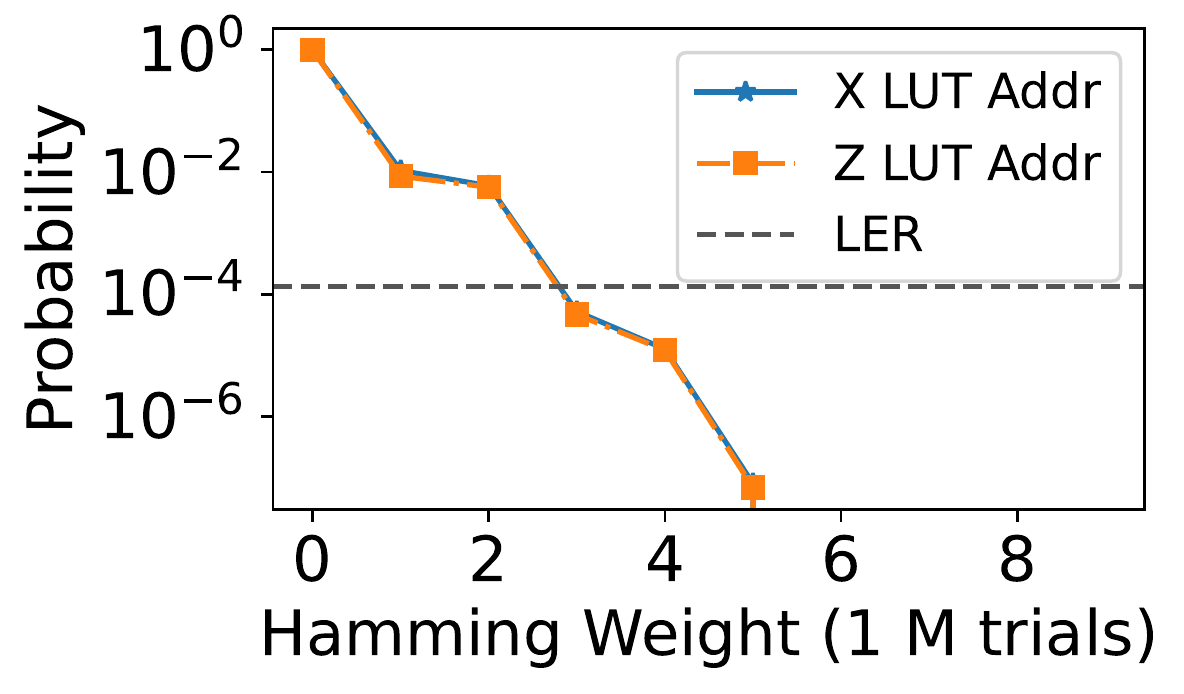} }}%
    \subfloat[\centering]{{\includegraphics[width=4.2cm]{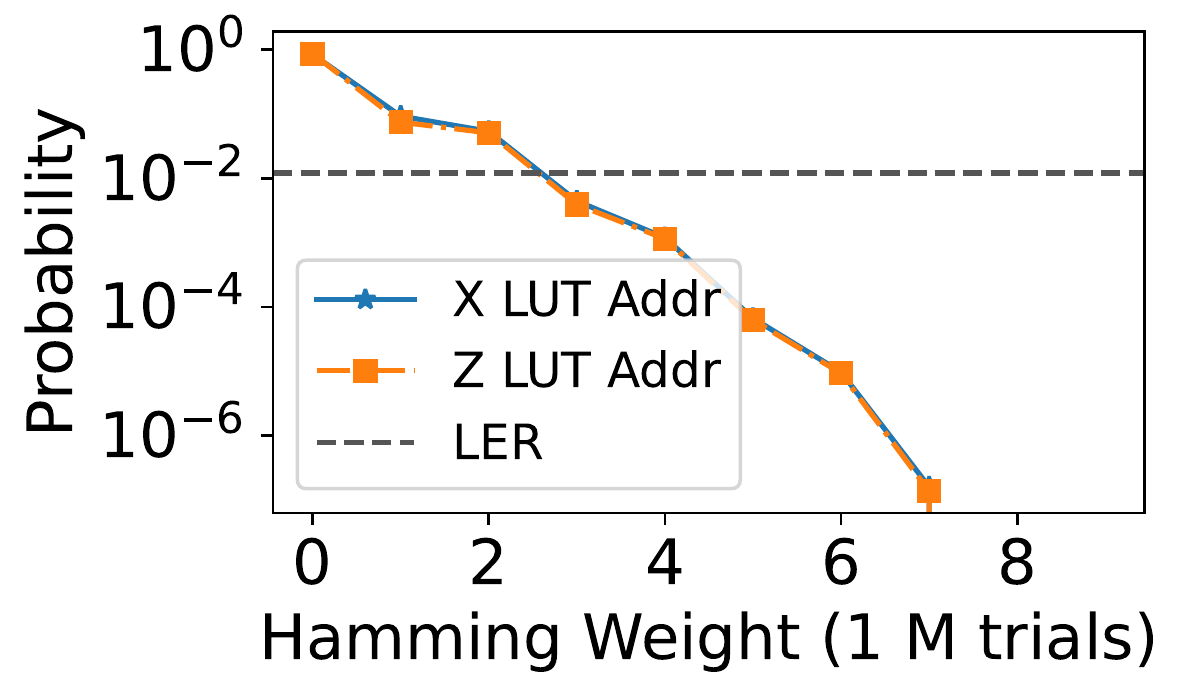} }}%
    \caption{Probability distribution of the Hamming Weight of memory addresses for (a) p = 0.1\% and (b) p=1\% for 5 cycles}%
    \label{fig:hammingweightdist3}%
\end{figure}

    \item As for small distances, a decoder can assign only a limited number of errors, the LUT entries contain a large number of zeros.
    \textit{Thus, the LUT entries themselves are compressible. }
    
\end{enumerate}
    \begin{tcolorbox}
    To summarize, not all LUT entries are accessed with equal probability and therefore, the size of the LUTs can be reduced by eliminating entries that are unlikely to be accessed. Furthermore, the data entries corresponding to the most probable error events that must be stored can be compressed too. 
    \end{tcolorbox}

\subsection{Compression in Software}
\label{sec:compression}
We investigate the scope of compression using the $[d=3, m=2]$ configuration. Figure~\ref{fig:hammingweightdist3} shows that the probability of accessing LUT addresses with 3 or more ones  is equal to or lower than the logical error rate. Even if these entries are not stored, the decoder accuracy is unlikely to be affected as the events would result in a logical error with low probability.

\begin{figure}[htb]
\centering
    \includegraphics[width=\columnwidth]{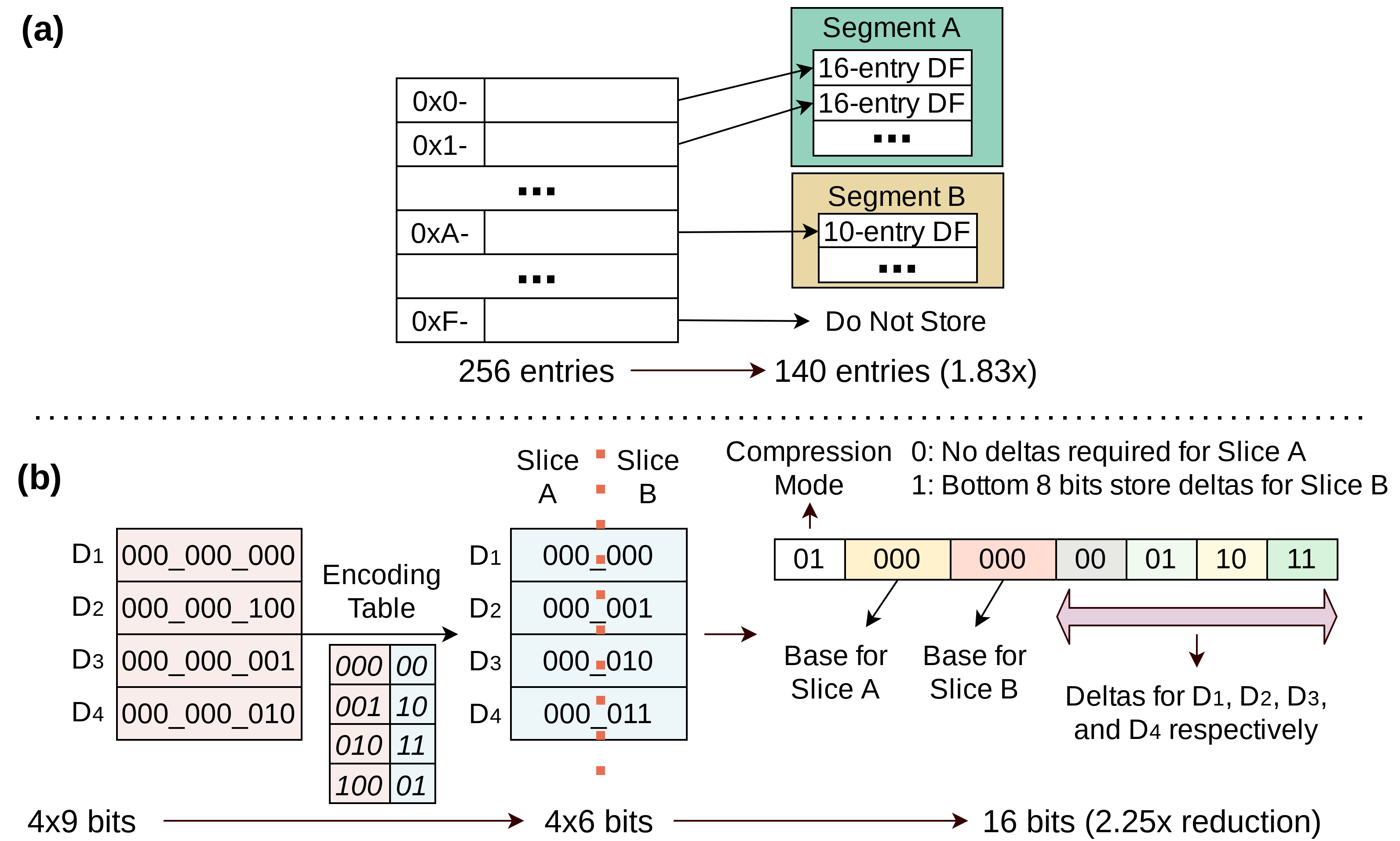}
    \caption{Steps (in software) involved in compressing LUTs.} 
    \label{fig:compression}
\end{figure}

To implement the CLUT, we split the address space into two groups-- Segments A and B, as shown in Figure~\ref{fig:compression}(a). These segments comprise of data frames (DFs)- a contiguous block of 16 and 10 entries respectively. A block of 16 consecutive LUT entries are assigned a 16-entry or 10-entry DF depending on the Hamming weight. Addresses with all zeros or a single one in the four most significant bits are assigned a 16-entry DF, whereas addresses with two ones in the four most significant bits are assigned a 10-entry DF. For example, as shown in Figure~\ref{fig:compression}(a), 0x00 to 0x0F correspond to a 16-entry DF in Segment A, whereas 0xA0 to 0xAA are assigned a 10-entry DF in Segment B. Addresses 0xAB to 0xAF are not stored as their Hamming weight is more than 3. Although we want to store LUT entries for addresses of Hamming weights up-to 3 only, as memory addresses are not contiguous in terms of Hamming weight, this results in memory fragmentation. Instead, we use data frames that cause some addresses with higher Hamming weights to be stored as well (such as 0x1F), but this ensures a simpler addressing scheme. This reduces the number of entries by 1.83x from 256 to 140. 

Next, the DF entries are compressed by re-encoding using the least number of bits and Base-Delta-Immediate~\cite{pekhimenko2012base} compression. For example, 9-bit error assignments from four consecutive entries are packed into a 16-bit word, as shown Figure~\ref{fig:compression}(b). First, each 9-bit data is encoded into 6-bit data using an encoding table. Next, four 6-bit entries are vertically sliced (Slice A and B) and packed into a 16-bit word. The 2-bit \textit{compression mode} is needed because other memory regions require packing slice A and B differently. For example, when all slice B bits are 0s and slice A is stored using a 3-bit base and four 2-bit deltas, the mode is $\textit{10}$. We investigated other compression schemes and found these to be most effective. Our studies show that for this configuration, compressing the internal state part of the DF entries needs a different scheme and the overheads of decompression exceed the benefits of compression as the data is only 4 bits. Finally, each CLUT reduces to 140 bytes, 3x lower than the full LUT (416 bytes).  As this is done in software, it does not incur hardware overheads.

\begin{figure}[htb]
\centering
    \includegraphics[width=\columnwidth]{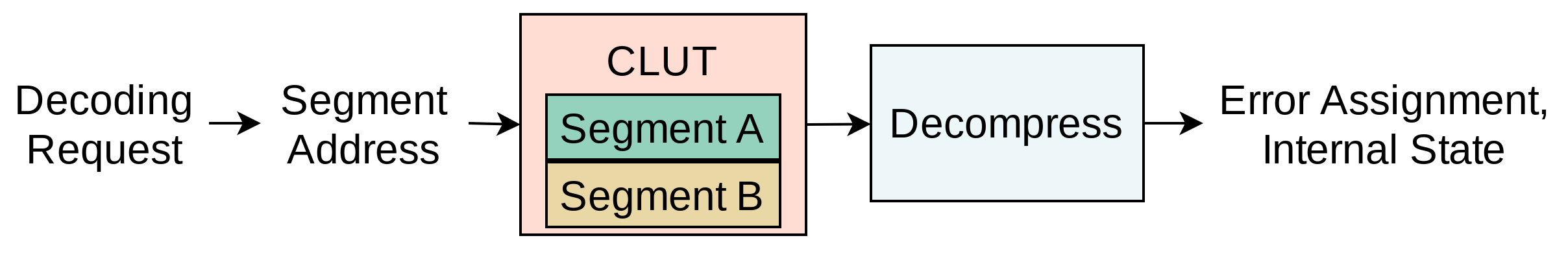}
    \caption{Steps involved in obtaining data from CLUTs.} 
    \label{fig:clutdatapath}
\end{figure}

\subsection{Accessing CLUTs and Decompression in Hardware}
Unlike the baseline design, the decoding request cannot be directly used to index into the LUT. In order to service a decoding request, it is routed to the appropriate CLUT segment and DF. The CLUT entry is decompressed to obtain the error assignments and internal state, as shown in Figure~\ref{fig:clutdatapath}. The hardware overhead to use CLUTs is the logic to obtain the segment address and perform decompression.

\subsection{Performance and Overheads of CLUTs}
\begin{figure}[htp]
    \centering
    \subfloat[\centering]{{\includegraphics[width=4.2cm]{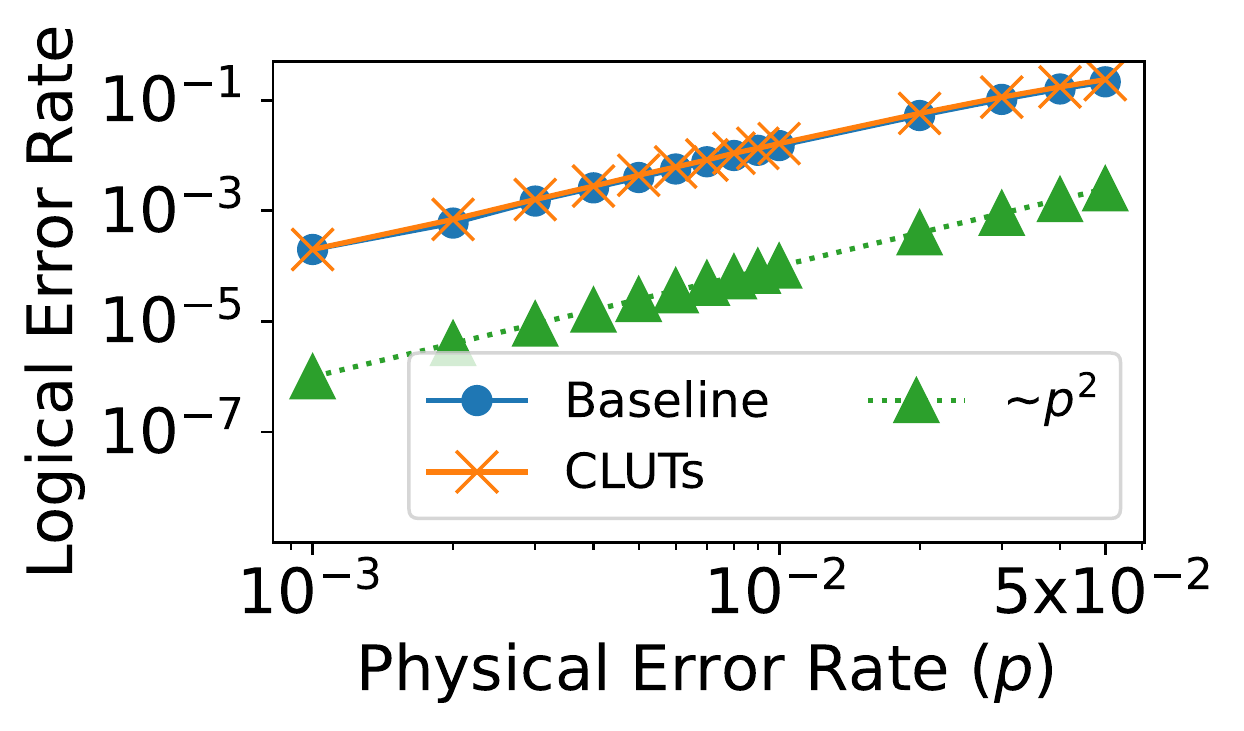} }}%
    \subfloat[\centering]{{\includegraphics[width=4.2cm]{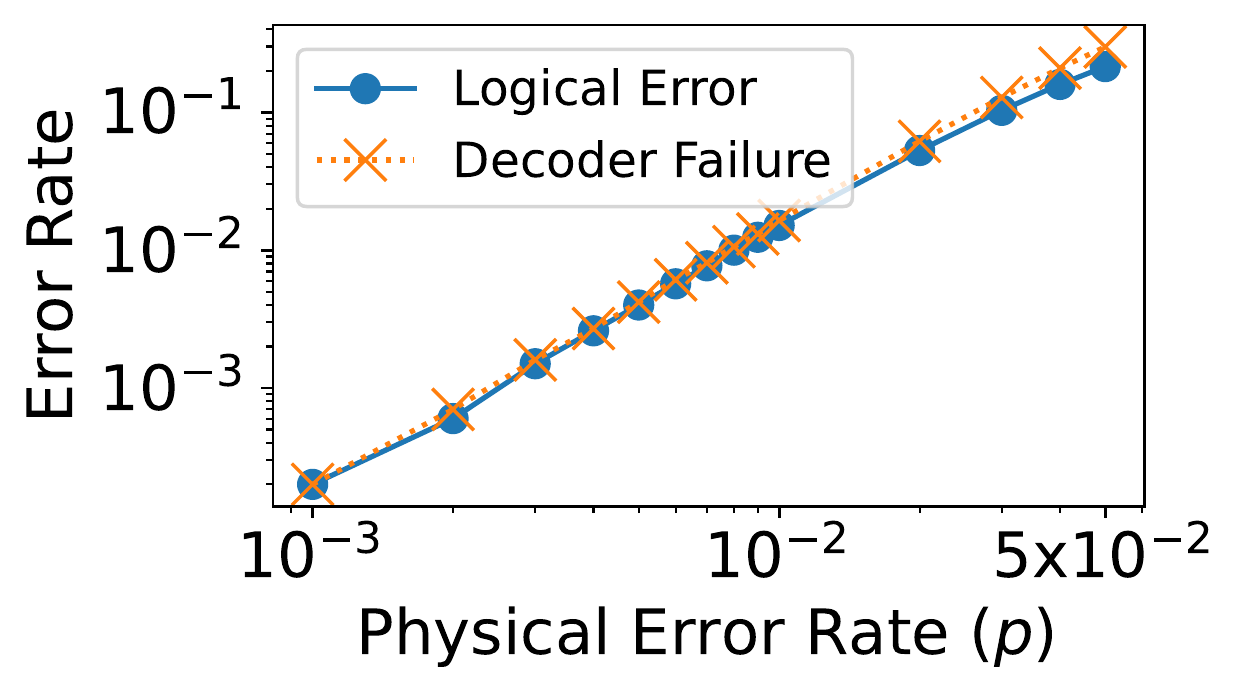} }}%
    \caption{(a) Comparison of logical error rate for the baseline and LILLIPUT with CLUTs (b) Logical error rate and decoder failure rate for LILLIPUT with CLUTs}%
    \label{fig:clutperformance}%
\end{figure}

Figure~\ref{fig:clutperformance}(a) compares the logical error rate of the baseline design with respect to LILLIPUT using CLUTs. We observe that CLUTs offer similar performance as the baseline and does not degrade the LER. Figure~\ref{fig:clutperformance}(b) shows the decoder failure rate (due to missing LUT entries) and we observe that using CLUTs does not lead to events that increase the failure rate. A decoder failure here refers to failure to decode because the LUT address requested is not present in the CLUT. 

Table~\ref{tab:logicutilizationclut} shows the logic overhead (increases from 6\% to 11\%) and memory reduction (by 3x) in using CLUTs on Cyclone FPGAs. On Arria V, both designs require less than 1\% logic utilization and thus, the overhead is acceptable.

\begin{table}[htb]
\begin{center}
\begin{small}
\caption{Logical Overhead for Implementing CLUTs}

\setlength{\tabcolsep}{1.2mm} 
\renewcommand{\arraystretch}{1.1}
\label{tab:logicutilizationclut}
{\footnotesize
\begin{tabular}{ |c|c|c|c|c|c|} 
\hline
\multicolumn{3}{|c|}{LILLIPUT (Baseline)} & \multicolumn{3}{c|}{LILLIPUT with CLUT} \\ 
\hline
LEs & Registers & Memory & LEs & Registers & Memory \\
\hline
\hline
353 (6\%) & 209 & 832 Bytes & 688 (11\%) & 261 & 280 Bytes \\

\hline
\end{tabular}}
\end{small}
\end{center}
\end{table}

\subsection{Scaling to Other Decoder Configurations}
\begin{figure}[htp]
    \centering    \subfloat[\centering]{{\includegraphics[width=4.2cm]{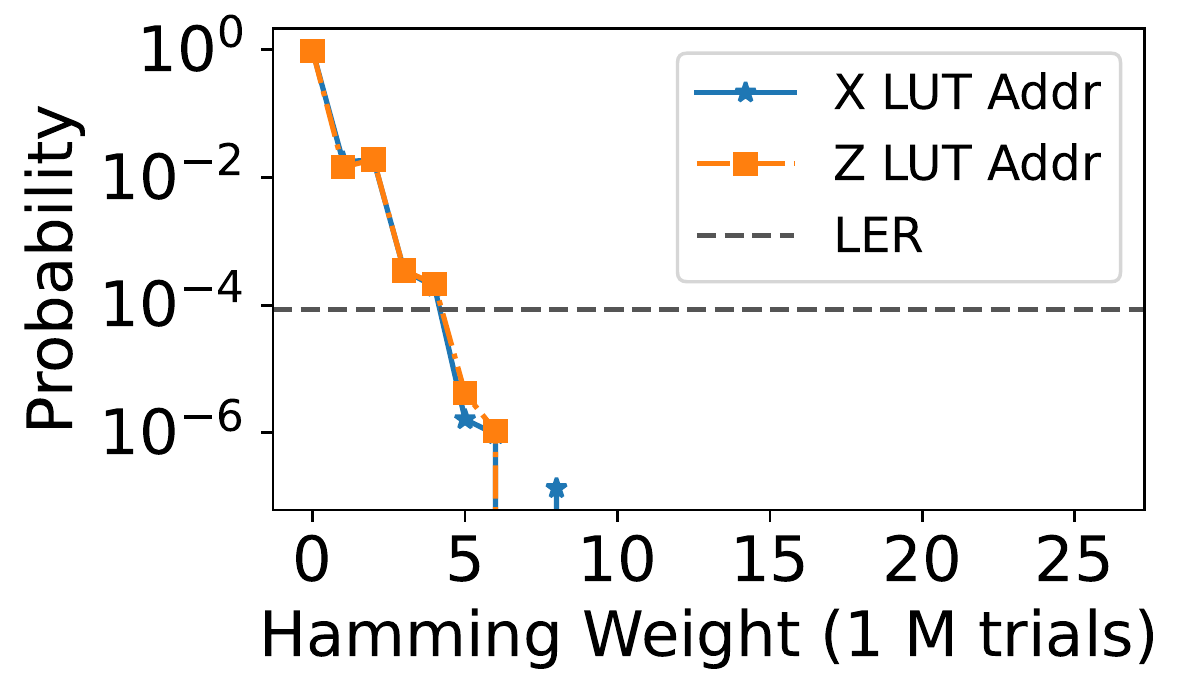} }}
    \subfloat[\centering]{{\includegraphics[width=4.2cm]{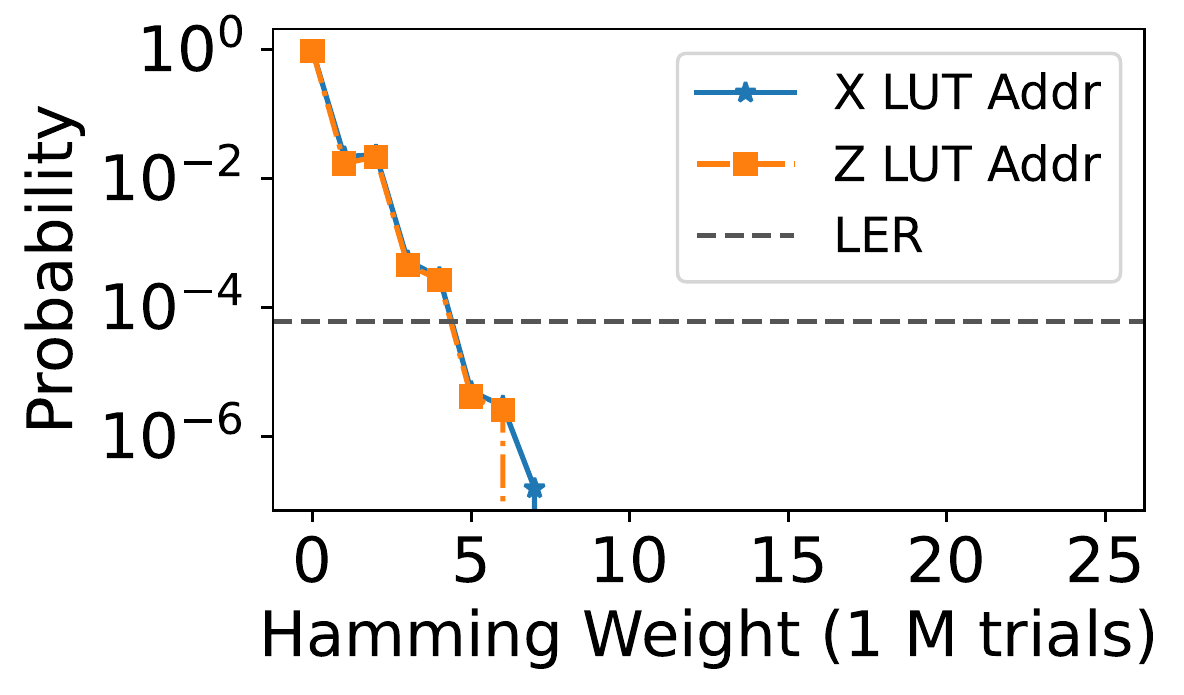} }}
    \caption{Probability distribution of Hamming weight of LUT accesses for (a) [d=4,m=3] and (b) [d=5,m=2] configurations.}
    \label{fig:hammingweightdist4}
\end{figure}

The LUTs for distance 4 and 5 surface codes can be reduced in a similar fashion. Figure~\ref{fig:hammingweightdist4} shows the probability distribution of the Hamming weight of the LUT addresses accessed for decoder configurations $[d=4,m=3]$ and $[d=5,m=2]$ respectively. We observe that storing only the LUT entries corresponding to Hamming weights 5 and below in the CLUTs are sufficient for both of these configurations. 

Table~\ref{tab:clutreduction} compares the size of CLUTs for distance 4 and 5 decoder configurations. We observe that using CLUTs can reduce the memory requirement by up-to 107x. The CLUTs fit within the embedded memory available on Arria II/V~\cite{intelarria} and Stratix 10 devices~\cite{intelstratix} and does not require any external memory access. Note that this estimate does not take into account compression of the DF entries in the CLUT (second step discussed in Section~\ref{sec:compression} as distances 4 and 5 require a different compression scheme and a detailed discussion is beyond the scope of this paper. To avoid the address translation for CLUT lookups, an alternative option is to use Cuckoo hashing~\cite{pagh2004cuckoo} which incurs 2x memory overhead, but still fits the budget of these FPGA boards. Lastly, to support LILLIPUT for higher decoder configurations $[d=5,m=3]$ and $[d=5,m=4]$, the LUTs can be compressed and accessed from an external memory.

\begin{table}[htb]
\begin{center}
\begin{small}
\caption{Memory Requirement for LILLIPUT with CLUTs}

\setlength{\tabcolsep}{1.2mm} 
\renewcommand{\arraystretch}{1.2}
\label{tab:clutreduction}
{\footnotesize
\begin{tabular}{ |c|c|c|c|c|} 
\hline
Decoder & \multirow{2}{*}{Design} & \multicolumn{3}{c|}{Memory (Baseline)} \\ 
\cline{3-5}
Configuration & & X & Z & Total \\
\hline
\hline
\multirow{2}{*}{$[d=4,m=3]$} & Baseline & 5.75 MB & 48 MB & 53.75 MB \\
\cline{2-5}
& W/ CLUT & 245 KB & 457 KB & 702 KB (78.4x)\\
\hline
\multirow{2}{*}{$[d=5,m=2]$} & Baseline & 74 MB & 74 MB & 148 MB \\
\cline{2-5}
& W/ CLUT & 704.6 KB & 704.6 KB & 1.38 MB (107x)\\
\hline
\end{tabular}}
\end{small}
\end{center}
\end{table}

%% file: sections/section8relatedwork.tex
\section{Related Work}
We discuss the related work on decoders and compression and compare as appropriate. 

\vspace{0.1in}
\noindent \textbf{Decoders for QEC}:
Developing accurate and fast decoders for QEC have been an area of research for several years. Recently, several hardware decoders~\cite{nisqplus,ueno2021qecool,das2020scalable} have been proposed for real-time decoding. However, these decoders have lower accuracy or rely on emerging technologies. In contrast, LILLIPUT is very accurate, requires a constant latency, is reconfigurable, and can be seamlessly integrated with existing quantum systems, making it an ideal candidate in the near future. Tomita et al. first investigated software LUT decoders using a fixed set of rules to determine the error assignments~\cite{tomita2014low}. However this design is only limited to distance 3 surface codes and 3 cycles. In contrast, LILLIPUT is an end-to-end solution that uses MWPM, accounts for device error models, and can be implemented over a wide number of configurations and QEC cycles.
Most recently, Ryan-Anderson et al. used software decoders to achieve real-time decoding for color codes on trapped-ion systems~\cite{ryan2021realization}. However, this design requires access to general-purpose CPUs with the associated overhead in time for transferring data. Unfortunately, superconducting qubits may not tolerate such large latencies because unlike trapped ions, superconducting qubits can retain information for only a few microseconds. LILLIPUT on the other hand is more versatile and can be adapted to both device technologies. In general, there is no widely accepted standard decoding algorithm for color codes similar to MWPM for surface codes~\cite{landahl2011fault,chamberland2020triangular}.  

\vspace{0.1in}
\noindent \textbf{Compression in conventional and quantum systems}:
Cache and memory compression in traditional architectures are mainly designed to optimize for capacity, bandwidth, and power. Their effectiveness varies depending on the sparsity of the data. Recently, Das et al. investigated syndrome compression for reducing the bandwidth needed for syndrome transmission in large quantum systems~\cite{das2020scalable} by exploiting the sparsity in syndrome data. In contrast, compression of LUTs in LILLIPUT relies on the fact that not all error events are equally likely and error assignments data tend to be sparse. In another work~\cite{wu2019full}, compression has also been found to be effective in simulating quantum circuits for program verification, a necessary step to evaluate quantum algorithms and machines.

%% file: sections/section9conclusion.tex
\section{Conclusion}
Demonstration of small surface codes using real-time decoding represents a significant milestone in quantum computing. In this paper, we propose LILLIPUT- a Lightweight Low Latency Look-Up Table decoder that is accurate, fast, and can be seamlessly integrated with near-term quantum hardware. LILLIPUT interfaces with the readout logic, detects error events and corrects them as they appear in real-time. LILLIPUT uses LUTs to identify errors instead of running a decoding algorithm. The LUTs are programmed offline using a software decoder and accounts for the error model of the quantum device. In this paper, we use minimum weight perfect matching and error rates that are representative of a Google Sycamore device~\cite{sycamoredatasheet}. LILLIPUT utilizes less than 7\% logic and requires up-to 148 MB of memory for distances up-to 5 on off-the-shelf FPGAs. To reduce the memory overhead of LILLIPUT, we propose Compressed LUTs (CLUTs). CLUTs exploit the fact that all error events are not equally likely and store information only for the most probable events. This reduces the memory required by up to 107x without deteriorating performance. 

\section*{Acknowledgement}
We would like to thank Kunal Arya, Michael Newman, Evan Jeffrey, Craig Gidney, and Austin Fowler for the technical discussions and feedback throughout the project.